\documentclass[11pt,preprint2]{aastex}

\newcommand{\slopefivewfpc}{$0.18 \pm 0.03$}
\newcommand{\slopesixwfpc}{$0.15 \pm 0.03$}
\newcommand{\slopefiveacs}{$0.15 \pm 0.02$}
\newcommand{\slopesixacs}{$0.20 \pm 0.06$}

\newcommand{\slopeJC}{$0.22 \pm 0.02$}

\begin{document}

\title{Tip of the Red Giant Branch Distances. II. Zero-Point Calibration}

\author{Luca Rizzi,\ R. Brent Tully,\ Dmitry Makarov\altaffilmark{1,2},\ and Lidia Makarova\altaffilmark{1,2}}
\affil{Institute for Astronomy, University of Hawaii, 2680 Woodlawn Drive,
 Honolulu, HI 96822}

\and

\author{Andrew E. Dolphin}
\affil{Steward Observatory, University of Arizona, Tucson, AZ~85721}

\and

\author{Shoko Sakai}
\affil{Division of Astronomy and Astrophysics, University of California, 
Los Angeles, CA~90095-1562}

\and

\author{Edward J. Shaya}
\affil{University of Maryland, Astronomy Department, College Park, MD 20743}

\altaffiltext{1}{also at Special Astrophysical Observatory of the Russian Academy of Sciences, 
 Nizhnij Arkhyz, 369167, Karachaevo-Cherkessia, Russia}
\altaffiltext{2}{Isaac Newton Institute of Chile, SAO Branch}

\begin{abstract}
The luminosity of the Tip of the Red Giant Branch (TRGB) provides an excellent measure of galaxy distances and is easily determined in the resolved images of nearby galaxies observed with Hubble Space Telescope. There is now a large amount of archival data relevant to the TRGB methodology and which offers comparisons with other distance estimators. Zero-point issues related to the TRGB distance scale are reviewed in this paper. Consideration is given to the metallicity dependence of the TRGB, the transformations between HST flight systems and Johnson-Cousins photometry, the absolute magnitude scale based on Horizontal Branch measurements, and the effects of reddening. The zero-point of the TRGB is established with a statistical accuracy of 1\%, modulo the uncertainty in the magnitude of the Horizontal Branch, with a typical rms uncertainty of 3\% in individual galaxy distances at high Galactic latitude. The zero-point is consistent with the Cepheids period-luminosity relation scale but invites reconsideration of the claimed metallicity dependence with that method. The maser distance to NGC 4258 is consistent with TRGB but presently has lower accuracy.
\end{abstract}

\section{Yet another calibration of the absolute magnitude of the TRGB}

The Tip of the Red Giant Branch (TRGB) is arguably  the most valuable distance indicator for galaxies within $\sim 10$ Mpc. It offers several advantages compared to other methods: (a) the observable is very bright, $M_I \sim -4$, (b) the physical processes behind the use of this method are well understood \citep{1997MNRAS.289..406S, 1997eds..proc..239M, 2002PASP..114..375S}, (c) the measurements are extremely efficient, considering it is possible to get a distance with only two observations, (d) it can be applied to almost all galaxy types requiring only that  they contain a significant old population, (e) it is technically easy compared to, e.g., a search for variable stars and determination of the pulsation parameters, and (f) RGB stars are dispersed so reddening in the host galaxy is not a significant problem.
On the other side, the TRGB is not a perfect standard candle. The bolometric magnitude of the tip is known to depend quite strongly on the metallicity of the underlying stellar population, and to a less extent on its age. Besides, the TRGB method is at the moment a tertiary distance indicator, calibrated on secondary distance indicators like RR Lyrae or Cepheids.

The first paper in this series \citep[][:Paper I]{2006AJ....132.2729M} discussed the mechanics of determining the TRGB.
The main purpose of the present paper is to critically review and verify the TRGB calibrations presented in the literature over the past years. Our primary goals are to: (a) provide a new estimate of the dependence of the TRGB magnitude on the color of the stellar population (i.e. the metallicity), (b) provide a new zero point of the TRGB vs. color relation and (c) provide a new calibration in the Hubble Space Telescope (HST) flight system, F555W, F606W, and F814W; both for the WFPC2 and the ACS cameras.

\section{The TRGB method}
Low mass, globular cluster-like stars begin their main sequence life by burning hydrogen in a radiative core surrounded by a convective envelope. When the hydrogen levels in the core drops, the energy production shifts to a shell that slowly moves outward. The star becomes redder and brighter, it approaches the Hayashi  line of fully convective objects, and gradually makes its way up the red giant branch (RGB). The helium core is degenerate and it keeps growing out of the ashes of the burning hydrogen. Its equation of state is independent of temperature. The core keeps getting warmer, until it reaches about 100 million K and triple-$\alpha$ helium burning can ignite. The ignition of a nuclear reaction in a degenerate core leads to a thermonuclear runaway, because the star cannot compensate the energy production with expansion, but in the case of the helium burning the degeneracy is removed before the star explodes. After this violent flash of energy, the stars rapidly contracts, and start to quietly burn helium in the core on the horizontal branch or red clump. Due to the presence of the degeneracy, the helium ignition happens at almost constant core mass. This in turn means that the ignition occurs at a predictable luminosity.

The observational evidence of this theoretical picture is a sharp cut-off of the luminosity function of the RGB, approximatively located at $M_I \sim -4$, probably first noted by W. Baade in \citeyear{1944ApJ...100..137B}. Almost 50 years after, the work of \cite{1990AJ....100..162D} and \cite{1993ApJ...417..553L} demonstrated the power of this stellar evolution phase in determining the distance to galaxies and globular clusters. Since then, a series of studies have been devoted to improving various aspects of the methodology. In particular, astronomers have been working in two directions, the first toward the absolute calibration of the observable, the magnitude of the tip of the RGB, the second toward the development of objective methods to estimate the position of the TRGB in the color-magnitude diagram (CMD).

By analyzing the CMD of a set of globular clusters, \cite{1990AJ....100..162D} derived the bolometric magnitude of the TRGB as:
$$ M_{bol}^{TRGB}=-3.81-0.19\rm{[Fe/H]}$$
which gives the magnitude in the $I$ band when combined with the bolometric correction:
$$ \rm{BC_I}=0.881-0.243(V-I)_0. $$
Their abundance calibration is based on the color of the RGB at $M_I = -3$, in the form:
$$ \rm{[Fe/H]}=-15.16 + 17.0 (V-I)_{0,-3}-4.9[(V-I)_{0,-3}]^2. $$ 
All these relations are defined for $-2.2 < \rm{[Fe/H]} < -0.7$ (M15, 47 Tuc).
The absolute scale is set on the luminosity of RR Lyrae stars using the calibration of  \cite{1990ApJ...350..155L}, $M_V(RR)=0.82 + 0.17 \rm{[Fe/H]}$. 
\citet{1993ApJ...417..553L} applied the method to the resolved stellar populations of nearby galaxies.
They demonstrated that the $I$ band luminosity function of the red giant branch is an excellent distance indicator because the metallicity dependence of the TRGB is much smaller in the $I$ band than it is in other optical bands, and because the sensitivity of common CCDs is well matched with observing red giants in the $I$ band.
They replaced the abundance calibration with a similar measurement, half a magnitude brighter:
$$ \rm{[Fe/H]}=-12.64 + 12.6 (V-I)_{0,-3.5}-3.3[(V-I)_{0,-3.5}]^2.$$
\citet{1993ApJ...417..553L} also introduced a reproducible and quantitative method to estimate the position of the TRGB (see later in this Section). 

In a series of papers \citep{1999AJ....118.1738F, 2000AJ....119.1282F, 2001ApJ...556..635B,
2004AA...424..199B}, these authors obtained a new robust calibration of the magnitude of the tip, extended to high metallicities (up to $\rm{[Fe/H]} = -0.2$) and to infrared passbands. For the $I$ band, they derive:
$$ M_I^{TRGB}=0.258 \rm{[M/H]}^2 + 0.676 \rm{[M/H]} -3.629, $$
where $\rm{[M/H]}$ is the global metallicity corrected for $\alpha$-enhancement.

\begin{figure}
\plotone{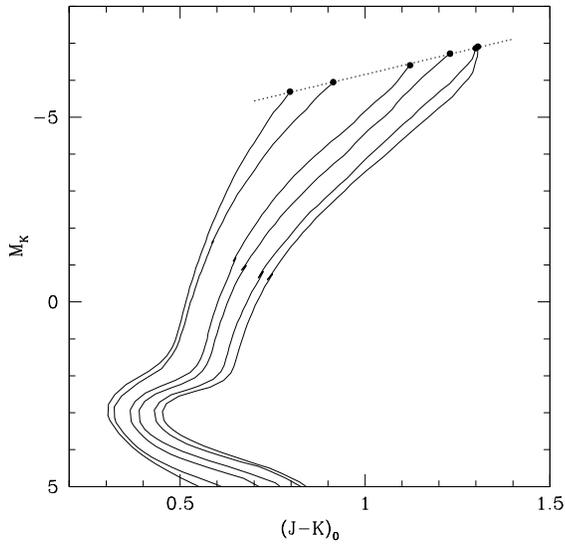}
\caption{Isochrones in IR passbands from the library of \citet{2002AA...391..195G}, with dots marking the position of the TRGB. The curves, from left to right, correspond to increasing values of metallicity; Z=0.004,0.001,0.004,0.008,0.19,0.30.}
\label{IR1}
\end{figure}

\begin{figure}
\plotone{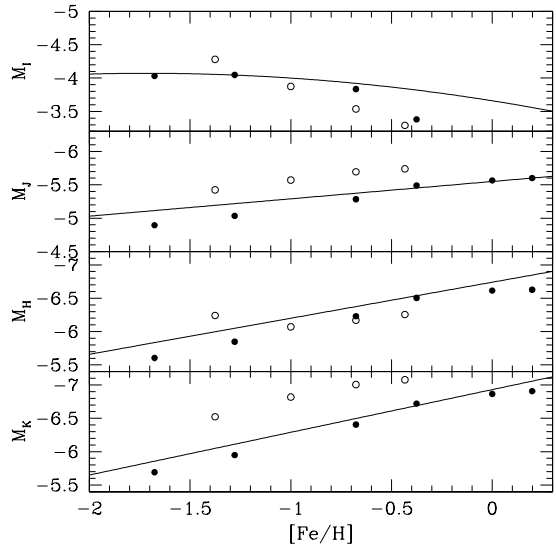}
\caption{Position of the TRGB vs. metallicity for I,J,H,K bands. Dark points refer to the isochrones from \citet{2002AA...391..195G}. Open points refer to the $\alpha$-enhanced models of \citet{2000AA...361.1023S}, and solid lines show the calibration of \citet{2004AA...424..199B}.}
\label{IR2}
\end{figure}

IR passbands offer a potentially valid alternative to the $I$ band, since the TRGB is brighter and the interstellar reddening is reduced. Most of the new generation telescopes will be optimized for observations in the near IR, including the JWST, pushing the possibilities of the TRGB method to much larger distances. On the other side, IR passbands have a few disadvantages that are shown in Figures \ref{IR1} and \ref{IR2}. 

Figure \ref{IR1} shows a series of isochrones from the library developed by the Padova stellar evolution group \citep{2002AA...391..195G}, with an improved thermally pulsing AGB as in \citet{2006AA...452..195C}, and for metallicities from $Z=0.004$ to $Z=0.30$. The dots show the position of the TRGB. 
Figure \ref{IR2} shows the passband dependence of the position of the TRGB vs. metallicity, using the same set of isochrones (filled dots). The open dots refer to the TRGB in $\alpha$-enhanced models 
\citep{2000AA...361.1023S}. The lines show the calibration of \citet{2004AA...424..199B}.
Figures \ref{IR1} and \ref{IR2} show that the metallicity dependence of the TRGB  is much higher in the IR passbands, which means that a safe application of the method for distance measurements requires a much more precise knowledge of the metallicity than is required for the $I$ band. Besides, the metallicity dependence is {\it positive}, meaning that the TRGB of metal rich stars is brighter than the TRGB of metal-poor stars. As a result, in a composite stellar population, with a large spread in metallicity/color, a clear detection of the tip becomes  more uncertain. It can be seen from Figure \ref{IR2}, that the condition of minimal metallicity dependency will be achieved with a passband around 1 micron. It will be interesting to make observations at this wavelength with the next generation of detectors in space.

There have been efforts to determine the metallicity dependence of the TRGB by means of stellar evolution theory \citep[see][for a review of work on the subject]{2002PASP..114..375S}.
In their review, the authors note that a fair agreement exists among the various predictions, with differences on $M_{bol}^{TRGB}$ of less than $0.1$ mag, notwithstanding the uncertainties affecting the equation of state of partially degenerate matter, neutrino energy losses, electron conduction opacity, and the nuclear cross section for the $3\alpha$ reaction. As an example, the relationship derived by
\citet{1998MNRAS.298..166S} is:
$$ M_{bol}^{TRGB} = -3.949 -0.178 \rm{[M/H]} +0.008 \rm{[M/H]}^2 $$

Note that all these calibrations (except for the theoretical ones) are based on measurement that do not meet the completeness criteria as set in  \citet{1995AJ....109.1645M}. In that paper, the authors demonstrate that a minimum amount of 100 stars in the first magnitude bin fainter than the tip are needed to safely detect the TRGB discontinuity. None of the globular clusters studied in \citet{1990AJ....100..162D} or \citet{1993ApJ...417..553L} meet this criterion. With their Galactic globular clusters, the number of stars in the first magnitude bin below the tip ranges from 2 to 20 and the adopted definition of the TRGB is the brightest and reddest star of the sample. A similar problem affects the measurements of \citet{1999AJ....118.1738F, 2000AJ....119.1282F}. The only case in which a sufficient number of stars has been used is presented in \citet{2001ApJ...556..635B}, where the pillar of the calibration is the CMD of $\omega$ Centauri. There is a need to provide a new calibration based entirely on the highly populated CMDs of resolved nearby galaxies, where no corrections for small sample is required.

In the early application of the method, the position of the TRGB was determined by eye inspection of the CMD or the luminosity function. \citet{1993ApJ...417..553L} introduced a more quantitative definition. They convolved the RGB luminosity function with a zero-sum Sobel kernel in the form $[-2,0,+2]$, and defined the TRGB as the peak of the filter response function. \citet{1996ApJ...461..713S} refined this method by replacing the histogram luminosity function with a continuous probability distribution, smoothed according to the photometric error. A different approach was suggested by \cite{2002AJ....124..213M}, who applied a maximum-likelihood technique based on  the assumption that the luminosity function at the TRGB can be described as a step function with a fixed logarithmic slope of 0.3 faintward of the the tip.

\begin{figure}[htbp]
\plotone{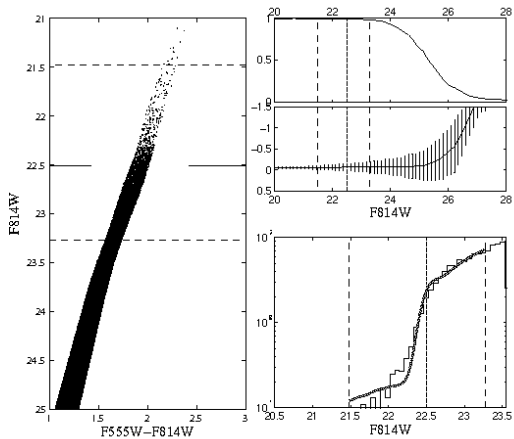}
\caption{Application of the TRGB detection method presented in \cite{2006AJ....132.2729M}. Left panel: Narrow slice of the CMD of NGC 300. Upper right panel: Completeness function and distribution of photometric errors. Lower right panel: Observed luminosity function (black) and best fitting model (red).\label{fit}}
\end{figure}

In this paper, we will use the method recently presented in Paper I. This method is an optimized version of the maximum likelihood approach that takes into account completeness, photometric errors, and biased error distributions. An example of the application of the method to a narrow color slice of the CMD of NGC 300 is presented in Figure \ref{fit} (the reasons for such a narrow color selection are discussed in Section \ref{cmds_coldep}). The left panel of Figure \ref{fit} shows the CMD, and the limits of the area selected for the fitting procedure. On the right half of the figure, the upper panel shows both the completeness function (on the top) and the distribution of photometric errors (on the bottom). The biased distribution of the errors is evident at magnitudes fainter than I=25.5. The lower panel shows the result of the fit: the observed luminosity function is indicated by the solid black line, while the red thick line shows the best fitting model. 

\section{Color dependence of the Tip of the Red Giant Branch.}
\label{cmds_coldep}
We can express the calibration equation of the absolute magnitude of the TRGB as:
$$M_{``I"}^{TRGB}= a + b (``V"-``I")$$ where $a$ is the zero-point and $b$ is the slope of the dependency on a particular choice of color ``V"-``I". In this equation, $``V"-``I"$ is the average color of stars at the tip of the giant branch (and not at a certain magnitude below the tip as in previous studies). 
The ``V" and ``I" passbands would traditionally be defined by the Johnson-Cousins V and I filters. The corresponding HST flight filters are F814W for ``I" and either F555W or F606W for ``V". The implications of these alternatives will be considered.

We first concentrate on the dependence of the TRGB on color, and for the moment we ignore the effect on zeropoints  that would arise from erroneous distances.
The sample of galaxies we have analyzed for metallicity effects includes: M33, NGC 300, NGC 1313, NGC 4258, NGC 4605, NGC 5128. Table \ref{tab_hst} identifies the HST archive observations used to derive the CMDs. All the images retrieved from the archive were reduced again using either {\sc hstphot} version 1.1 \citep{2000PASP..112.1383D}, for WFPC2 images, or Dolphot (http://purcell.as.arizona.edu/dolphot/) for ACS images. The same set of reduction parameter was used for all the datasets, and artificial star experiments were performed to estimate the completeness levels and the distribution of photometric errors. 

\begin{deluxetable}{lclccccl}
\rotate
\tablecaption{Source HST program for the metallicity calibration sample. \label{tab_hst}}
\tablewidth{0pt}
\tablehead{
\colhead{Galaxy} & \colhead{Program-Cycle} & \colhead{P.I.}  & \colhead{Instrument} & \colhead{Filters} & \colhead{$E(B-V)^{1}$} & \colhead{Slope} & \colhead{Adopted distance modulus}}
\startdata
M33 & 5914-5 & Sarajedini & WFPC2 & F555W,F814W & 0.042 & 0.22 $\pm$ 0.03 & $24.67 \pm 0.08$ \citep{2006AJ....132.1361S}\\
& 6640-6 & Mighell & WFPC2 & F555W,F814W & 0.042 & 0.21 $\pm$ 0.08 & \\
& 8059-7 & Casertano &  WFPC2 & F606W,F814W & 0.042 & 0.20 $\pm$ 0.11& \\
NGC 5128 & 8195-8 & Harris &  WFPC2 & F606W,F814W & 0.110 & 0.21 $\pm$ 0.07& $27.87 \pm 0.16$ \citep{2004AA...413..903R} \\
NGC 300 & 9492-11 & Bresolin &  ACS & F555W,F814W & 0.013 & 0.22 $\pm$ 0.04& $26.63 \pm 0.06$  \citep{2004ApJ...608...42S}\\
NGC 4258 & 9477-11 & Madore &  ACS & F555W,F814W & 0.016 & 0.22 $\pm$ 0.03 & $29.47 \pm 0.09 \pm 0.15$ \citep{2001ApJ...553..562N}\\
NGC 4605 & 9771-12 & Karachentsev &  ACS & F606W,F814W & 0.014 & 0.21 $\pm$ 0.10 & $28.69 \pm 0.17$ \citep{2006AJ....131.1361K}\\
NGC 1313 & 10210-13 & Tully &  ACS & F606W,F814W & 0.109 & 0.20 $\pm$ 0.07 & $28.21 \pm 0.20$ (this work)\\
\enddata
\tablenotetext{1}{From \cite{1998ApJ...500..525S}}
\end{deluxetable}

Reddening corrections for all the galaxies of the sample were taken from the dust maps of 
\citet{1998ApJ...500..525S} and are summarized in column 6 of Table \ref{tab_hst}. In many cases, accurate reddening determinations are available as a result of photometric studies, but we have preferred to maintain a uniform approach to avoid the subtleties of having to decide galaxy by galaxy which value is the most reliable. For example, reddening estimates based on Cepheid studies tend to suggest values systematically higher than the simple foreground values \citep[e.g., see][]{2005ApJ...628..695G}, but these values may only be relevant to the young and possibly dust-enshrouded Cepheids and not to the whole galaxy. We return to the problem of reddening in a later section.

The CMD of each galaxy of the sample was divided in color slices, roughly following the shape of globular clusters ridge lines. The width of the slices was decided independently for each galaxy, taking into account the number of stars in the tip region, the quality of the photometry, and the color range covered by the RGB. For each CMD strip, we determined the position of the TRGB, and the average color of the stars at the tip. Figures from \ref{m33_1} to \ref{ngc4258} in Appendix \ref{cmds} show the CMD of the galaxies of the sample. Only the region close to the TRGB is shown. Dark points near the TRGB show the actual measurements.  Solid lines serve as  graphic guides to the slope of the TRGB as a function of colors.

\subsection{Slope of the $M_I \, {\rm vs.}\, (V-I)$ relation in the HST flight filters}
\label{slope}


We now proceed to provide a measure of the slope of the relation between $M_{F814W}$, the absolute magnitude of the TRGB in the F814W filter, and the color defined either as $F555W-F814W$ or $F606W-F814W$, both for the WFPC2, and the ACS detectors.

\subsubsection{WFPC2 flight system}

\begin{figure}
\plotone{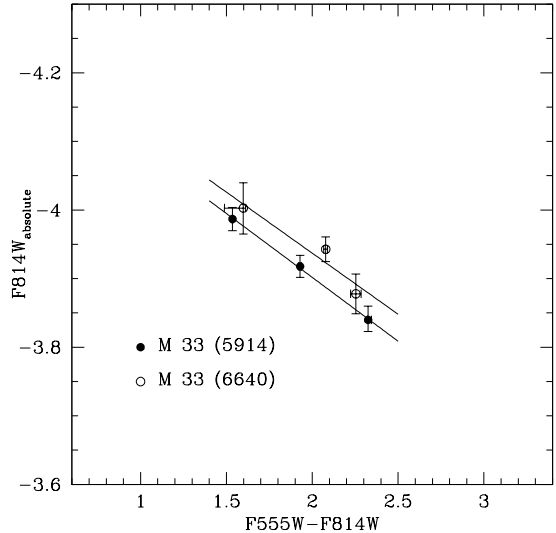}
\caption{Magnitude of the TRGB in the F814W filter as a function of (F555W-F814W) color, in the WFPC2  photometric system. Solid lines are the best fits to the points.\label{slope555_wfpc2}}
\end{figure}

\begin{figure}
\plotone{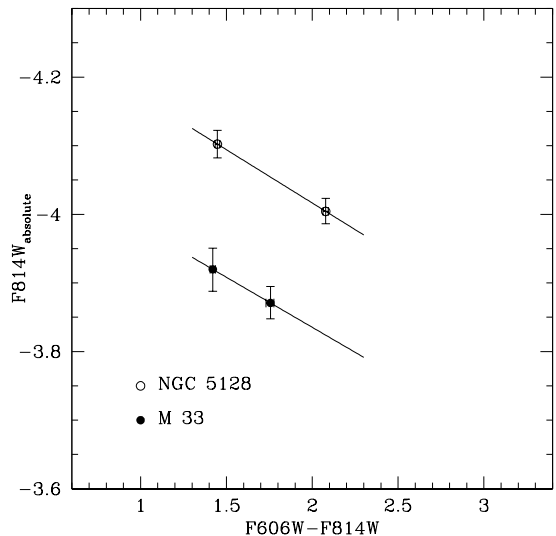}
\caption{Magnitude of the TRGB in the F814W filter as a function of (F606W-F814W) color, in the WFPC2 photometric system. Solid lines are the best fits to the points.\label{slope606_wfpc2}}
\end{figure}

M33 is the only galaxy in our sample that was observed with WFPC2 with the F555W filter. The data come from two HST programs, 5914 and 6640, and the corresponding CMD have been measured independently. For the F606W filter we have data both for M33 and NGC 5128 (Centaurus A). The measurements have been translated to the absolute plane by using the following distance moduli:
for M 33, $(m-M)_0=24.67$ \citep{2006AJ....132.1361S} ; for NGC 5128, $(m-M)_0=27.87$ \citep{2004AA...413..903R}. The use of moduli from the literature allows us to compare results for different galaxies conveniently but it will be seen that we do not need highly accurate input distances. Figure \ref{slope555_wfpc2} shows the color dependence in the $(M_{F814W},F555W-F814W)$ plane with these data, while Figure \ref{slope606_wfpc2} shows the color dependence in the $(M_{F814W},F606W-F814W)$ plane. The slopes averaged over the separate fields are:

\begin{itemize}
\item $(M_{F814W},F555W-F814W)$: \slopefivewfpc
\item $(M_{F814W},F606W-F814W)$: \slopesixwfpc
\end{itemize}
There are evident zero point displacements. The issue of the zero points  will be discussed in Section \ref{absolute}.

\subsubsection{ACS flight system}

The other four galaxies of our metallicity sample were observed with ACS, namely NGC 300, NGC 4258, NGC 4605, and NGC 1313. NGC 300 and NGC 4258 were observed in the F555W filter, while NGC 4605 and NGC 1313 were observed in the F606W filter. The measurements have been translated to the absolute plane by using the following distance moduli: for NGC 300 $(m-M)_0=26.63$ \citep{2004ApJ...608...42S}; for NGC 4258 $(m-M)_0=29.47$ \citep{2001ApJ...553..562N}; for NGC 4605  $(m-M)_0=28.69$ \citep{2006AJ....131.1361K}. For NGC 1313 we have adopted a distance of $(m-M)_0=28.21$ based on the analysis in this paper.

\begin{figure}
\plotone{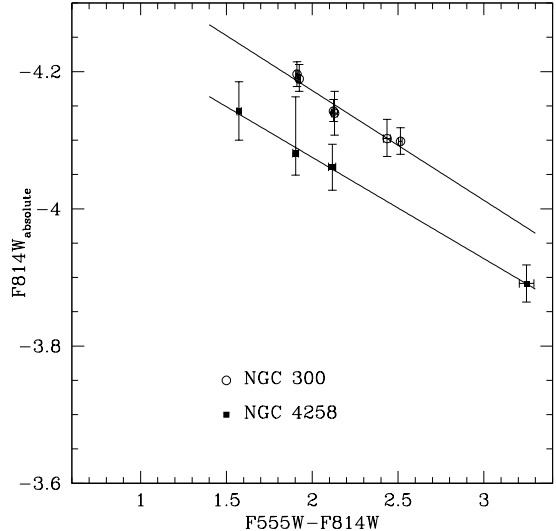}
\caption{Magnitude of the TRGB in the F814W filter as a function of (F555W-F814W) color in the  ACS photometric system.  Solid lines are the best fits to the points.  \label{slope555_acs}}
\end{figure}

\begin{figure}
\plotone{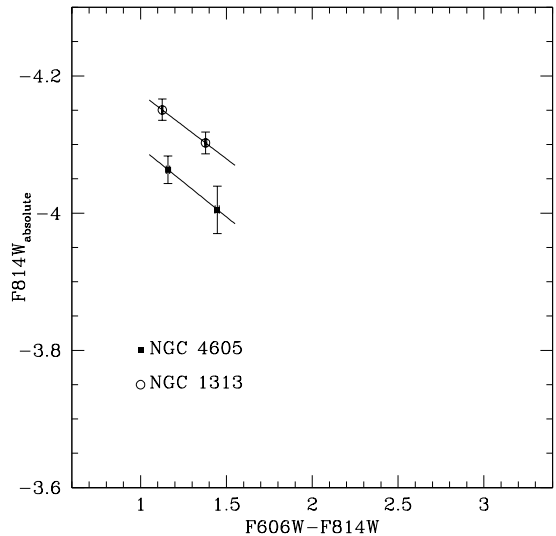}
\caption{Magnitude of the TRGB in the F814W filter as a function of (F606W-F814W) color in the ACS photometric system.  Solid lines are the best fits to the points.  \label{slope606_acs}}
\end{figure}

The ACS results are shown in Figures \ref{slope555_acs} and \ref{slope606_acs}, for the $(M_{F814W},F555W-F814W)$ and the $(M_{F814W},F606W-F814W)$  CMD, respectively.

The averaged slopes we measured are:

\begin{itemize}
\item $(M_{F814W},F555W-F814W)$: \slopefiveacs
\item $(M_{F814W},F606W-F814W)$: \slopesixacs
\end{itemize}
The zero point displacement of these results will be discussed in Section \ref{absolute}.

\subsection{Slope of the $M_I\, {\rm vs.} \, (V-I)$ relation in the Johnson-Cousins photometric system.}
\label{JC}
The slopes derived in the previous Section provide information necessary to derive distances to galaxies using the TRGB method that relies completely on an HST flight system, thus eliminating the uncertainties related to the photometric conversion to a ground system. However, there are multiple variations of the HST flight system. The Johnson-Cousins photometric system is more familiar, and it is still the only system for which a direct conversion between color and metallicity is available for RGB stars. For these reasons we convert our measurements from the HST flight systems to the Johnson-Cousins system.

Rather than convert the photometry of the single stars in a CMD, repeat the color selection, and derive new TRGB measurements, we directly convert the final TRGB measurements. Each of the TRGB measurements in our sample can be treated as a single representative star. The magnitude of this star is the magnitude of the TRGB in that specific color slice, and the color is the average color of the stars at the level of the tip. 

\begin{figure}
\plotone{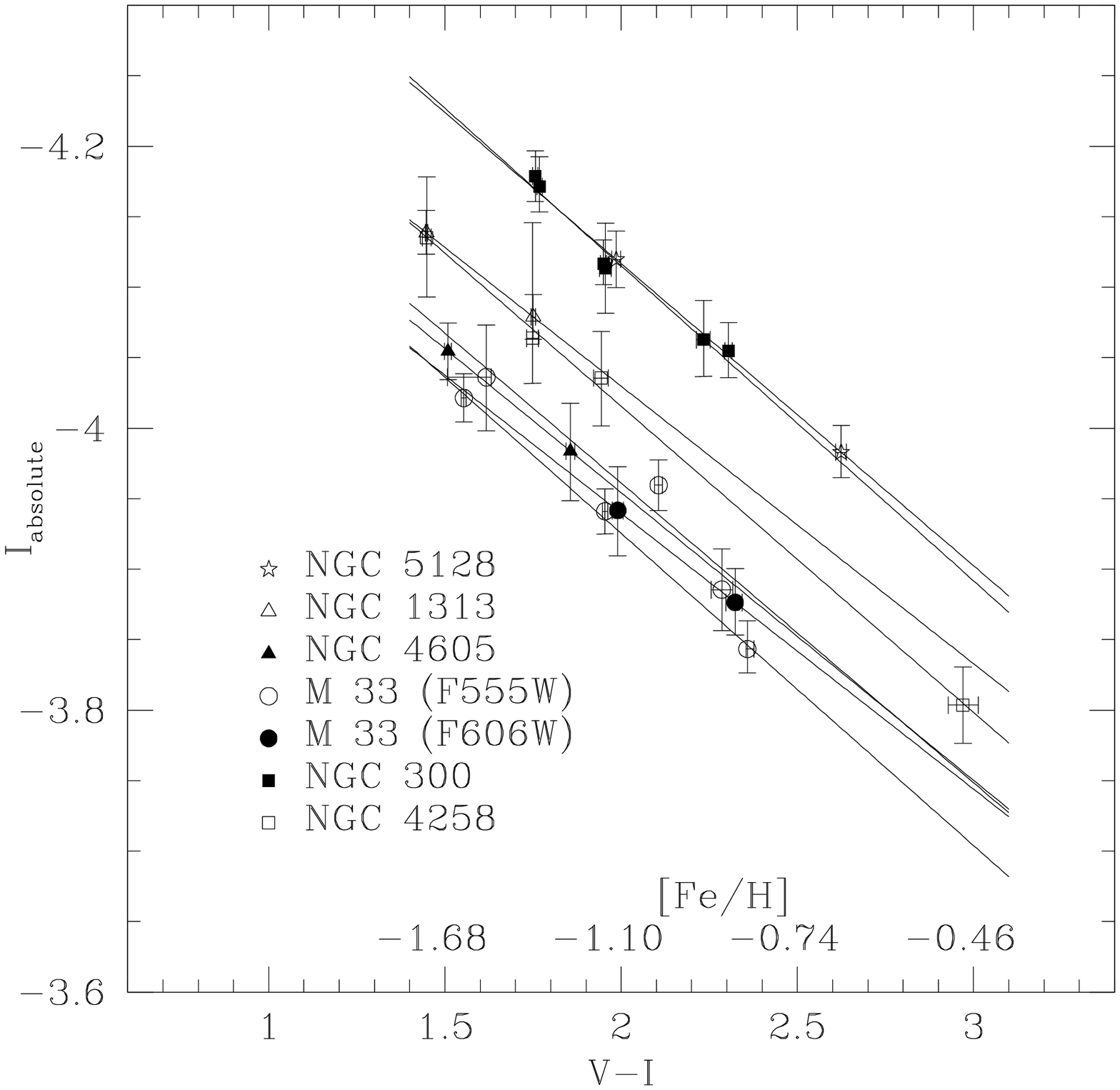}
\caption{Magnitude of the TRGB in the I filter as a function of  (V-I) color in the Johnson/Cousins photometric system. \label{trgb_JC}}
\end{figure}

Conversion from the HST flight system to the Johnson-Cousins system has been performed using the recipes presented in \citet{2005PASP..117.1049S} for ACS, and those in \citet{2000PASP..112.1397D} and \citet{1995PASP..107.1065H} for WFPC2. The results are shown in Figure \ref{trgb_JC} and the measurements for each galaxy are presented in column 7 of Table \ref{tab_hst}.
The weighted average of the slopes is \slopeJC.


\section{Constraining the zero-point}
\label{absolute}
While the slopes for the cases shown in Figure \ref{trgb_JC} are consistent, the zero-points are clearly incompatible. The vertical scatter must arise from inconsistent input distances.

Constraining the zero-point of the relations we have derived requires a rather complicated sequence of steps, and a series of assumptions on the physical properties of the galaxies of the sample. Figure \ref{trgb_JC} teaches us that a good calibration of the TRGB zero point requires exquisitely good distances to the calibrators.

Several methods exist to estimate the distance to resolved galaxies, such as those based on the use of Cepheid and RR Lyrae variable stars, the horizontal branch (HB), main sequence fitting, and red clump. In some special cases, even geometric methods can be applied, such as those that interpret the motion of water masers and binary stars. We will not discuss here the advantages and disadvantages of all the different methods, but only comment on the two most widely used: Cepheids variables, for Population I stars, and RR Lyrae/HB for Population II stars. Cepheids variable are very bright, and most of the galaxies of our sample have been successfully searched for these objects. Unfortunately distances derived with this method cannot be used to calibrate TRGB distances because the metallicity dependence of the period-luminosity relation is calibrated using TRGB measurements \citep{2004ApJ...608...42S}. We would end up in a circular argument. Instead we leave an absolute comparison with the Cepheid scale to the end, and use it only as a compatibility check.

\begin{figure}
\plotone{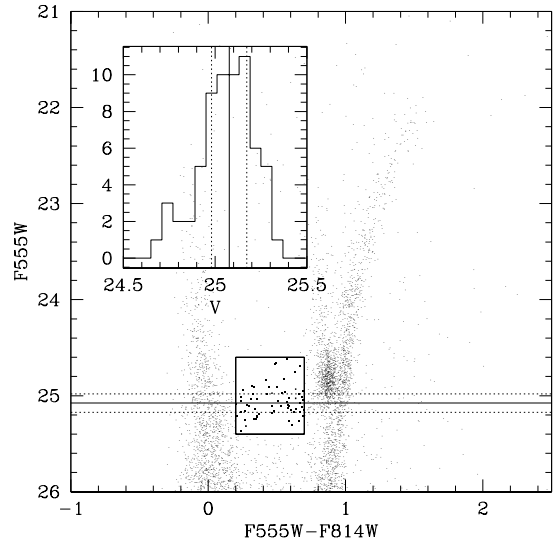}
\caption{Measurement of the HB level in IC1613, in the HST flight system. In the main panel, the square box shows the selection we used to define HB stars. The horizontal continuous line indicates the position of the HB, and the two dashed line show the rms of the measurement. The subpanel shows the histogram of the stars selected for the measuement. The vertical continuous and dashed lines have the same meaning as in the main panel.  \label{ic1613_hb}}
\end{figure}

\begin{figure}
\plotone{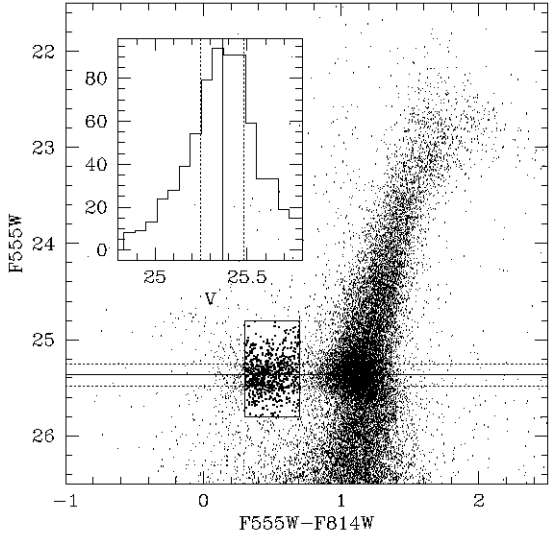}
\caption{Measurement of the HB level in NGC 185, in the HST flight system \label{ngc185_hb}}
\end{figure}

\begin{figure}
\plotone{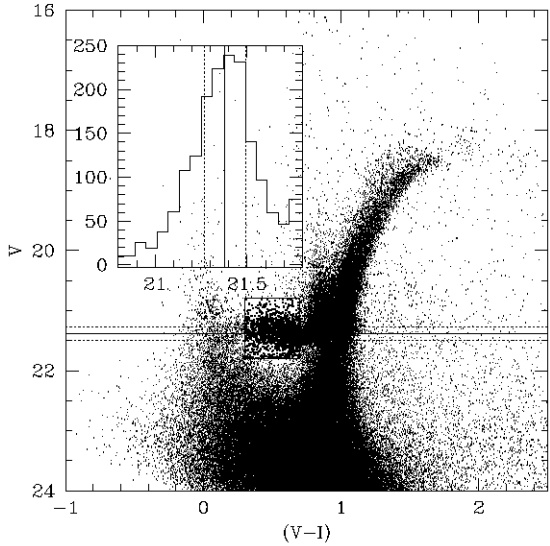}
\caption{Measurement of the HB level in the Fornax dwarf spheroidal. \label{fornax_hb}}
\end{figure}

\begin{figure}
\plotone{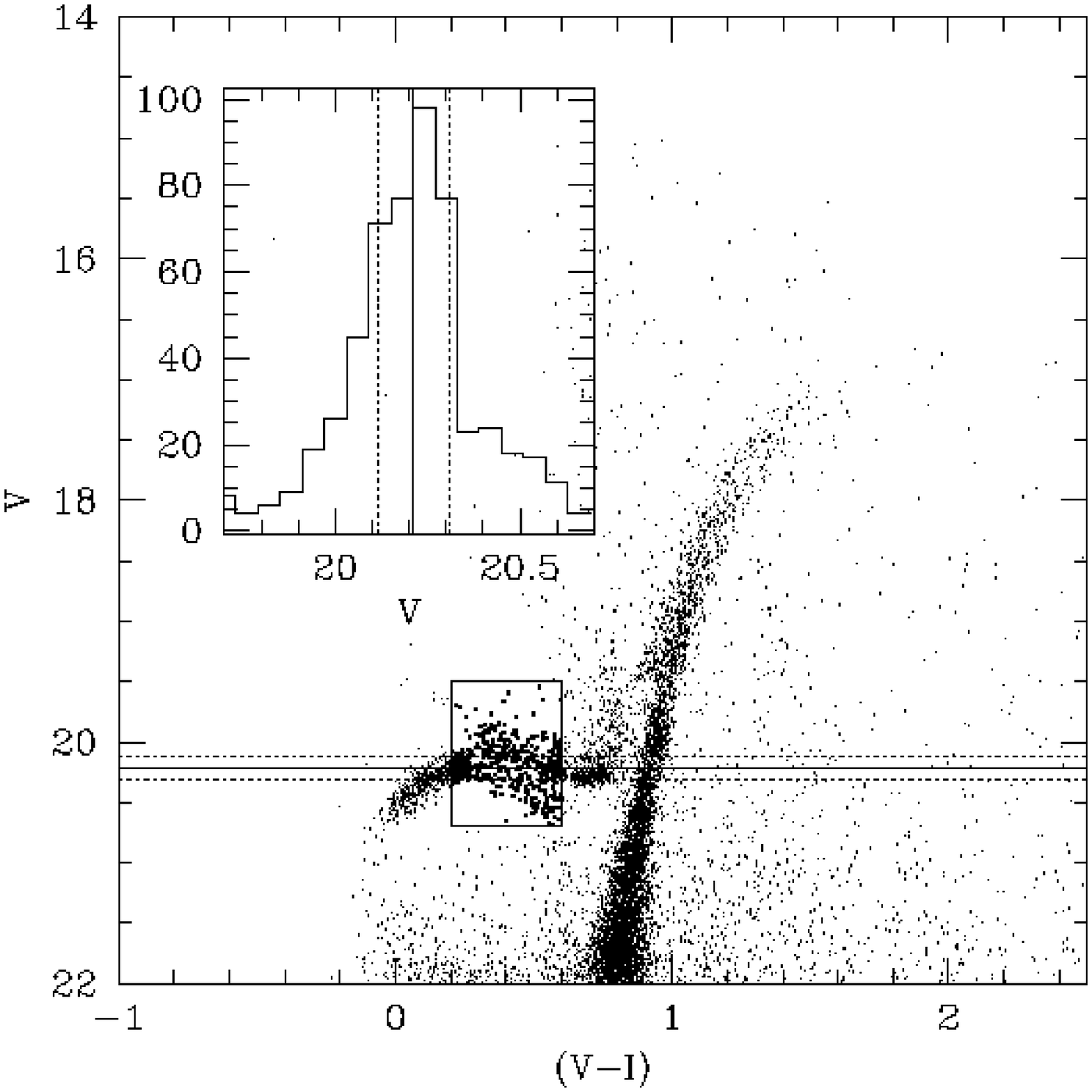}
\caption{Measurement of the HB level in the Sculptor dwarf spheroidal. \label{sculptor_hb}}
\end{figure}

A good alternative to a calibration with Cepheids variables is offered by HB and RR Lyrae stars. To have full control on the internal consistency of our measurements, we prefer to limit our analysis to galaxies for which an HST or ground based CMD exist that reaches well below the HB, and perform a new reduction. We did not give attention to RR Lyrae stars, but instead decided to use HB stars. Studies of RR Lyrae stars in the galaxies of the Local Group do exist, but in many cases they are quite old and based on photographic material.  The final list of good candidates include: IC 1613, NGC 185, the Sculptor and Fornax dwarf spheroidals, and M 33. Deep HST images are available for each of these galaxies, and archive material from the ESO 2.2m telescope was used for Sculptor and Fornax. 
We selected stars within $\sim$ 1 mag of the HB, and with $0.2 < V-I < 0.6$, and measured the average magnitude both in the HST flight systems, where available, and in the Johnson-Cousins system.
Figures \ref{ic1613_hb} through \ref{sculptor_hb} show the selection of stars used, the histogram of the selected stars, and the level of the measured HB, for IC 1613, NGC 185, Fornax and Sculptor, respectively. For M33, we used the measurements presented in \cite{2006AJ....132.1361S}. For the absolute magnitude of HB stars we used the calibration presented in \cite{2000ApJ...533..215C}
$$ M_V(HB) = (0.13 \pm 0.09)(\rm{[Fe/H]}+1.5) + (0.54 \pm 0.04)$$
The metallicity is on the scale of \citet{1997AAS..121...95C} in this relation.
We prefer to adopt this calibration because it directly provides the magnitude of the HB, rather than the magnitude of RR Lyrae stars.
The values we measured are presented in Table \ref{tab_vhb}.

\begin{deluxetable}{lcccccccccc}
\rotate
\tablewidth{0pt}
\tablecaption{Horizontal branch and TRGB magnitudes.\label{tab_vhb}}
\tablehead{
\colhead{Galaxy} & \colhead{Instrument}  & \colhead{$V_{HB}$} & \colhead{Std. dev.} & \colhead{$I_{TRGB}$} & \colhead{$(V-I)_{TRGB}$} & \colhead{E(B-V)} & \colhead{Adopted [Fe/H]\tablenotemark{1}}  & \colhead{$M_I$\tablenotemark{2}}}
\startdata
IC 1613 & WFPC2 &  25.06 $\pm$ 0.10 & 0.003 & $20.37 \pm 0.04$ & $1.60 \pm 0.02$ & 0.025 & -1.28 & -4.057\\
NGC 185 & WFPC2 &  25.35 $\pm$ 0.11 & 0.004 & $20.43 \pm 0.02$ & $2.01 \pm 0.04$ & 0.182  & -1.02 & -4.051\\
M33 & ACS & 25.92 $\pm$ 0.05 & \nodata & $20.69 \pm 0.02$ & $1.66 \pm 0.01$ & 0.042 & \nodata & -4.047\\
Fornax & WFI@2.2 &  21.37 $\pm$ 0.10 & 0.003 & $16.75 \pm 0.03$ & $1.64 \pm 0.03$ & 0.020 & -1.50 & -4.019\\
Sculptor & WFI@2.2 &  20.21 $\pm$ 0.10 & 0.004 & $15.60 \pm 0.03$ & $1.50 \pm 0.03$ & 0.018 & -1.74 & -4.050\\
\enddata
\tablenotetext{1}{On the \citet{1997AAS..121...95C} scale.}
\tablenotetext{2}{As plotted in Figure \ref{final_calib}.}
\end{deluxetable}

%

%








The major source of uncertainty in the distance determination is related to the estimate  of the metal  content of the galaxies of the sample. Numerous values exist in the literature for the metallicities of these objects, mostly photometrically derived, and it is often the case that there is no consensus on the average values, and the spread is significant. To maintain internal consistency, and to provide a reproducible result, we decided to measure the metallicity using the color of the giant branch at the level of the tip, using the calibration provided by \citet{2001ApJ...556..635B}. They found: 
$$(V-I)_0^{TRGB}=0.581 \rm{[Fe/H]}^2 +2.472 \rm{[Fe/H]} +4.013$$
which can be inverted to derive metallicities from the color of the stars at the TRGB. In this case, the resulting metallicity is on the \cite{1984ApJS...55...45Z} scale, and it needs to be converted into the \citet{1997AAS..121...95C} scale in order to use it to derive the absolute magnitude of the HB.

\begin{figure}
\plotone{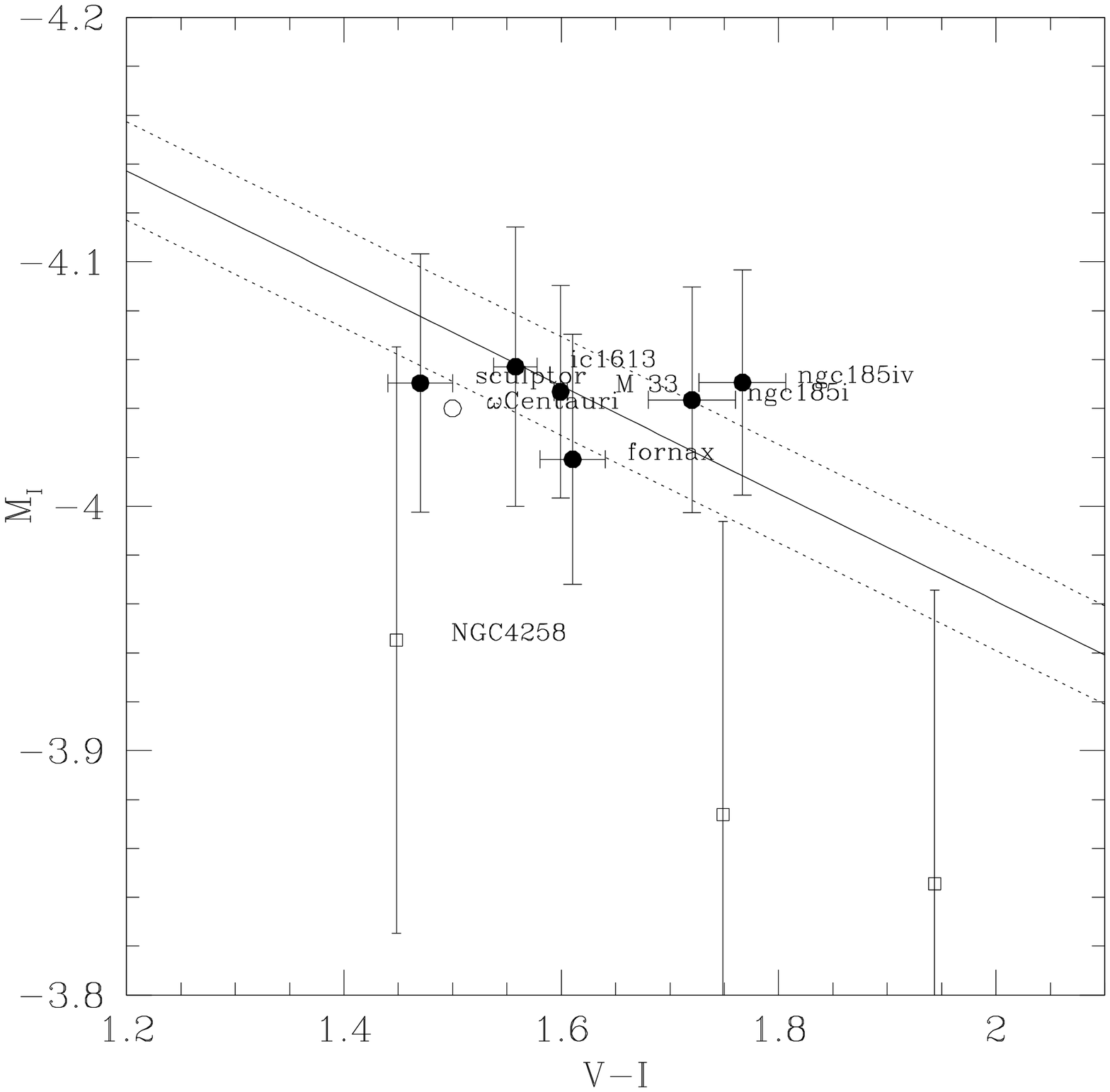}
\caption{Zero point calibration of the TRGB dependence on color. \label{final_calib}}
\end{figure}

Finally, the TRGB was measured for each of the galaxies, using the same method adopted in previous sections, and both the $V_{HB}$ and the TRGB values were adjusted from the HST flight systems as necessary and corrected for interstellar absorption using the values from \citet{1998ApJ...500..525S}. The final result is shown in Figure \ref{final_calib}. A fixed-slope (0.22/mag) fit to the data yields a value of the $M_I^{TRGB}$ at $(V-I)=1.6$ ([Fe/H] $\sim$ -1.5):
$$M_I^{TRGB}=-4.05 \pm 0.02.$$ This particular color (metallicity) value has been chosen as representative of the average color of the TRGB in the galaxies of the sample. The determination of the zero-point determined using $\omega$ Centauri by \citet{2001ApJ...556..635B} is shown with an open dot. The error was computed by propagating all sources of errors through the procedure followed to derive the value of $M_I^{TRGB}$. In particular, we first transformed the error in the color of the TRGB into an error in metallicity, and used the error in metallicity to derive the error in $M_V(HB)$. This error was finally added in quadrature to the error in the TRGB determination and to the error in the magnitude of the HB to obtain the vertical bars displayed in Figure \ref{final_calib}. We then performed an independent fixed-slope fit to the maximum and minimum values of $M_I$ for the different galaxies (upper and lower ends of the vertical error bars), and used the average result as our estimate of $M_I$. Half of the difference between the two determinations, divided by the square root of the number of points, was assumed to be the final error.
To derive the zeropoint for the HST flight system, again we have used the conversion relation presented in  \citet{2005PASP..117.1049S} for ACS, and in \citet{2000PASP..112.1397D} and \citet{1995PASP..107.1065H} for WFPC2, applied to a representative star having $(V-I)=1.6$, and $M_I=-4.05$.  The conversion of the magnitudes of this star into the HST flight system allows us to derive the following relations:
$$ M_{F814W}^{ACS}=-4.06+0.15[(F555W-F814W)-1.74]$$
$$ M_{F814W}^{ACS}=-4.06+0.20[(F606W-F814W)-1.23]$$
$$ M_{F814W}^{WFPC2}=-4.01+0.18[(F555W-F814W)-1.58]$$
$$ M_{F814W}^{WFPC2}=-4.01+0.15[(F606W-F814W)-1.12]$$
$$ M_I^{JC}=-4.05 + 0.217[(V-I)-1.6] $$ 
With this zeropoint, we have equivalent calibration relations for the various HST flight systems and the Johnson-Cousins system. 

The calibration that we would derive using the recent maser distance to NGC 4258 is indicated as well \citep{2005ASPC..340..466H}. The three open squares correspond to three different CMD color slices. Our calibration is consistent within the $1 \sigma$ uncertainty of the maser distance. However the 0.14 mag offset of NGC 4258 is a $2.3 \sigma$ departure from the TRGB distance.

\section{Reddening uncertainties}
With the small internal errors that are being claimed for the TRGB methodology, it is easy to appreciate that reddening uncertainties play a crucial role in accurately determining distances. In some cases, reddening variations across the galaxies can be large enough to prevent a meaningful measurement of the distance if only an average reddening is adopted. An example of this situation is provided by NGC 6822, a galaxy with a large angular extent that lies towards the galactic center at the low Galactic latitude $b=-18$. Four different HST WFPC2 fields from HST program 8314 result in very different determinations of the luminosity of the TRGB. The fields are identified as c1, c12, c18 and c25, the names referring to star clusters in this dwarf galaxy. Fields c18, c12, seem to have similar reddening, yielding TRGB luminosities  $I=19.93$ and $I=19.97$, respectively, but the TRGB luminosity of field c1 is $I=20.08$, while for c25 we find $I=19.86$. There is a range of 0.22 mags from the brightest to the faintest.   The average reddening measured on the maps of \cite{1998ApJ...500..525S} is $E(B-V)=0.236$. Clearly, an average value is not adequate to describe the complicated reddening distribution to the foreground of this galaxy.  

There are features of the CMD that provide a measure of reddening. If present, the blue main sequence probably provides the best handle. Other possibilities are the lower part of the RGB (less sensitive to age and metallicity than the upper part) and the red clump. The CMD of NGC 6822 is deep enough to allow us to test the effectiveness of all three features in determining the local reddening. Although a comparison with either isochrones or synthetic stellar populations is possible, we chose to compare directly with a galaxy with a very small reddening and similar stellar populations, to avoid the subtleties of choosing the right isochrone set or correctly applying observational errors to artificial stellar populations. A good candidate for the comparison is IC 1613, with an average reddening of $E(B-V)=0.025$. 

\begin{figure}[HT]
\plotone{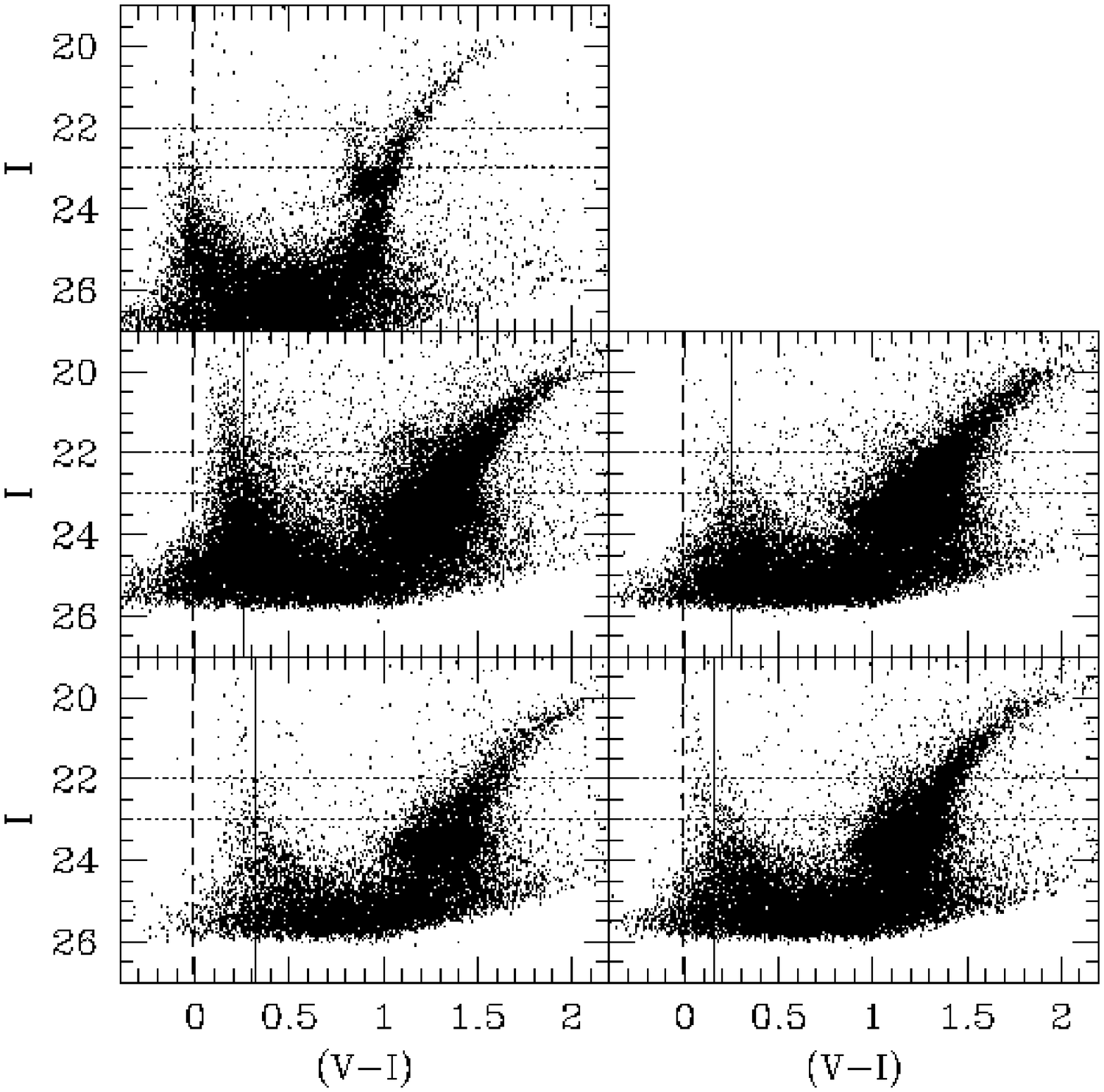}
\caption{Color-magnitude diagrams of IC 1613 (top left; with magnitude shift to match NGC 6822 distance) and of four fields in NGC 6822. Vertical lines show the average color of stars on the main sequence, selected within the region indicated by the horizontal dashed lines. Vertical dashed line marks the average color of the same feature in IC 1613. \label{fig:reddening}}
\end{figure}

The procedure we adopted is as follows. The CMD of IC 1613 and of the four fields in NGC 6822 were shifted in luminosity so that the tip (not reddening corrected) was at $M_I=-4$. The average colors of stars in the main sequence, lower red giant branch, and red clump were then measured for IC 1613 and for each field in NGC 6822, and the color differences between the corresponding features were used to determine the relative reddening difference between NGC 6822 and IC 1613. The small reddening of IC 1613 was finally added to the measurements. 
Figure \ref{fig:reddening} illustrates this procedure applied to main sequence stars. The upper left panel shows the CMD of IC 1613 (shifted in magnitude to match the distance of NGC 6822), while the other four panels refer to fields c12, c18, c25, and c1 respectively, in clockwise order from upper left. The average color of main sequence stars within the selection region (marked by the horizontal dashed lines) is indicated by the vertical solid lines, while the position of the same feature in IC 1613 is indicated by the vertical dashed line. It is evident that different fields show different color displacements, attributable to the intrinsic reddening of the particular field.

Results  of the main sequence comparison are shown in Table \ref{tab_reddening}. Columns 1 and 2 identify the galaxy and the field, column 3 records the TRGB luminosity in the F814W filter, column 4 gives the local reddening of the field from the main sequence offset, and finally column 5 shows the reddening corrected F814W luminosity of the TRGB. Very similar results were found using the red clump and the lower part of the red giant branch. In the case of the red clump, a noticeable difference in shape exists between IC 1613 and NGC 6822, so we adopted a selection region large enough to account for this effect. The reddening deduced from the position of the main sequence range from $E(B-V)=0.139$ to $E(B-V)=0.254$. Once the different reddenings are applied, the scatter in the measured luminosity of the TRGB is only 0.04 mags.

\begin{deluxetable}{lcccc}
\tablewidth{0pt}
\tablecaption{Determination of the differential reddening in NGC6822 \label{tab_reddening}}
\tablehead{
\colhead{Galaxy} & \colhead{Field} & \colhead{TRGB(F814W)} & \colhead{$E(B-V)$} & \colhead{TRGB(F814W,Corrected)}}
\startdata
IC1613 & \nodata & 20.44 & 0.025 & 20.34 \\
NGC6822 & c1 &  20.08 & 0.254 & 19.54 \\ 
NGC6822 & c12 & 19.97 & 0.203 & 19.53 \\
NGC6822 & c18 & 19.93 & 0.200 & 19.50 \\
NGC6822 & c25 & 19.86 & 0.139 & 19.54 \\
\enddata
\end{deluxetable}

\section{Comparison with the Cepheid distance scale}

In this section we compare distances obtained with the new calibration of the TRGB absolute magnitude with results obtained using Cepheid variables. 
\cite{2004ApJ...608...42S} (S04) recently provided a comparison between TRGB and Cepheids distances for a sample of nearby galaxies. They used WFPC2 to obtain V and I images of 7 galaxies, and supplemented the sample with an additional 10 galaxies for which TRGB and Cepheid distances were already available in the literature. With the intention of revisiting the issue, we retrieved all the observations mentioned in S04 from the HST archive, and performed a new uniform reduction using the latest version of HSTPhot 
\cite{2000PASP..112.1383D}.   After reductions, we measured the tip using the method presented in \cite{2006AJ....132.2729M}. Reddening values were taken from the maps of \cite{1998ApJ...500..525S}. Metal content measurements were taken from  S04. Results are presented in Table \ref{tab_cepheids}. The galaxy name is in column 1, followed by the I-band luminosity of the tip measured in this work (column 2) and in S04 (column 3), respectively. Column 4 contains the reddening as measured from the dust maps of \cite{1998ApJ...500..525S}. Finally, columns 5, 6, and 7 contain distance determinations, obtained with the TRGB method in this work (column 5), with the TRGB method in S04 (column 6), and with Cepheid variables in S04 (column 7). In this last case, S04 reports several different estimates of the Cepheid-based distances, using different calibrations and different photometric bands. The one reported here is from column 5 of Table 3 in S04, and is based on data in all photometric bands and the calibration published in \cite{1991PASP..103..933M}. There is no specific reason to adopt this version of the Cepheid-based distances, but we have verified that using any other of the distances published in S04 does not change the results.

\begin{deluxetable}{lccccccc}
\rotate
\tablewidth{0pt}
\tablecaption{Comparison of TRGB and Cepheids PLR results. \label{tab_cepheids}}
\tablehead{
\colhead{Galaxy} & \colhead{$I_{TRGB}^{This work}$} & \colhead{$I_{TRGB}^{S04}$} & \colhead{$A_I$}   & \colhead{$12+{\rm log(O/H)}$} & \colhead{$(m-M)_{0,TRGB}^{This work}$} & \colhead{$(m-M)_{0,TRGB}^{S04}$} & \colhead{$(m-M)_0^{Cepheids}$}}
\startdata
Sextans A & 21.78 & 21.73 & 0.090 & 7.49 & 25.78 & 25.67 & 25.85 \\ 
NGC 598 & 20.91 & 20.95 & 0.080 & 8.82 & 24.71 & 24.81 & 24.56 \\ 
NGC 3031 & 23.91 & 24.13 & 0.160 & 8.75 & 27.70 & 28.03 & 27.75 \\ 
NGC 3351 & 25.92 & 26.54 & 0.050 & 9.24 & 29.92 & 30.39 & 30.03 \\ 
NGC 3621 & 25.38 & 25.45 & 0.160 & 8.75 & 29.26 & 29.36 & 29.21 \\ 
NGC 5457 & 25.32 & 25.40 & 0.020 & 9.20 & 29.34 & 29.42 & 29.21 \\ 
IC 1613 & 20.29 & 20.25 & 0.050 & 7.86 & 24.38 & 24.31 & 24.29 \\ 
IC 4182 & 24.17 & 24.20 & 0.030 & 8.40 & 28.23 & 28.25 & 28.36 \\ 
WLM & 20.92 & 20.83 & 0.070 & 7.74 & 24.93 & 24.77 & 24.92 \\ 
Sextans B & 21.76 & 21.60 & 0.060 & 7.56 & 25.79 & 25.63 & 25.63 \\ 
NGC 3109 & 21.62 & 21.63 & 0.130 & 8.06 & 25.57 & 25.52 & 25.56 \\ 
LMC & 14.54 & 14.54 & 0.000\tablenotemark{1} & 8.50 & 18.57 & 18.59 & 18.50 \\ 
SMC & 14.95 & 14.95 & 0.000\tablenotemark{1} & 7.98 & 18.98 & 18.99 & 18.99 \\ 
NGC 224 & 20.53 & 20.53 & 0.150 & 8.98 & 24.37 & 24.47 & 24.41 \\ 
NGC 300 & 22.54 & 22.62 & 0.030 & 8.35 & 26.48 & 26.65 & 26.63 \\ 
\enddata
\tablenotetext{1}{Original source only provides dereddened values.}
\end{deluxetable}

The average difference between Cepheids-based minus our TRGB-based distance moduli is measured to be $-0.01 \pm 0.03$. The error is the standard deviation with 15 cases and an rms scatter of 0.10 mag per case. It is to be noted that we find a very good agreement between the Cepheids-based and TRGB-based scales along the whole range of metallicities, {\it without} introducing a correction for metallicity dependence of the Cepheid PLR. This is an unexpected result, and we prefer to postpone its discussion to a companion paper.

\cite{2006astro.ph..8211M} recently used ACS observations of NGC 4258 to provide a new estimate of the distance to this galaxy based on the Cepheid PLR, and found a distance modulus $\Delta(m-M)_0=10.88 \pm 0.04 \pm 0.05$ relative to the LMC, corresponding to $(m-M)_0=29.38$ if we assume $(m-M)_0^{LMC}=18.5$. To further investigate the relation between TRGB and Cepheids PLR distances, we performed a new reduction of the \cite{2006astro.ph..8211M} data set, and derived a new estimate of the luminosity of the TRGB. The data set consists of two fields, an inner one located close to the nucleus, and an outer one. \cite{2006astro.ph..8211M} derived the TRGB luminosity in the outer field, and found $I_{TRGB}=25.42 \pm 0.02$.  We measured the TRGB luminosity in both the inner and outer fields, obtaining $I_{TRGB}=25.45$ and $I_{TRGB}=25.52$, respectively. We assume that the difference between the two measurements is entirely due to measurements errors, and that the true value of the TRGB luminosity is the average of the two. The final distance modulus we derive is $(m-M)_0=29.42 \pm 0.06$, in very good agreement with the Cepheids distance scale, if $(m-M)_0^{LMC}=18.5$ is assumed.



\section{Conclusions}
At this point in time, the TRGB method appears to provide the best way to derive distances to relatively nearby galaxies. The intrinsic accuracy per object is as good as the best of alternatives. Single orbit HST/ACS observations can comfortably make a TRGB detection out to $\sim$ 10 Mpc. The procedure works with a feature of old stellar populations, which almost every nearby galaxy is found to possess, and this feature is not hard to isolate from younger populations. It can be entertained that accurate, methodologically coherent, distances could be obtained for all $\sim$ 500 unobscured galaxies within $\sim$ 10 Mpc.

In Paper I, as a first step towards the realization of this goal, \cite{2006AJ....132.2729M} refined procedures for the isolation of the TRGB with a maximum likelihood analysis. If the TRGB is situated well above the sensitivity limit of the observations then the TRGB is easily identified in a substantial population of old stars. However,  the identification of the TRGB location is subject to biases if near to the photometric limit. The maximum likelihood procedure informed by artificial star recovery statistics was found to overcome bias problems. The main result of that first paper in this series was the establishment of a recipe for the reproducible and unbiased identification of the luminosity marker we call the TRGB.

With this second paper we address issues concerning the intrinsic luminosity of the TRGB. We consider the following items:
\begin{enumerate}
\item The bolometric luminosity of the TRGB is strongly dependent on metallicity and even in our preferred I band there is a weak metallicity dependency. The slope of this dependency could be unambiguously identified within individual galaxies exhibiting a large metallicity range. Details of the dependency were traced in WFPC2 and ACS data separately and F555W-F814W or F606W-F814W colors, as summarized in Table \ref{tab_hst}, and related back to the Johnson-Cousins photometric system. An important characteristic of the methodology at I band is the slightly negative dependence of TRGB luminosity with metallicity. All observed galaxies have low metallicity old stars. Descending in luminosity, the onset of the TRGB is seen first with this characteristic old, low metallicity population.
\item The zero-point of the TRGB relation required a new calibration. We want to avoid linkage with the Cepheid scale except by way of comparison at the end. Besides, the zero-point of the Cepheids scale is none too firm, generally being set by the distance of the Large Magellanic Cloud alone. Here we establish the TRGB scale via an assumed luminosity for the Horizontal Branch and the identification of this feature in five Local Group galaxies (IC 1613, NGC 185, Sculptor, Fornax, M33). We arrive at the calibration
$$ M_I^{JC}=-4.05 (\pm 0.02) + 0.22 (\pm 0.01) [(V-I)-1.6] $$ demonstrated in Figure \ref{final_calib}
and the variations in the HST flight system given in Section \ref{absolute}.
\item Uncertainties in reddening become a dominant source of error. Relative to most other methods the problem is not serious because observations are made in the near IR and of an old population that can be found in minimally affected regions. Mostly, the problem is foreground contamination for which there are estimates \citep{1998ApJ...500..525S}. Independently, we have demonstrated that it is possible to estimate reddening from the same CMD that gives the TRGB measure as long as there is a well formed main sequence. 
\end{enumerate}

It is found that our zero-point is in fine agreement with the Cepheids scale for 15 comparison objects ($\mu_{Ceph}-\mu_{TRGB}=-0.01 \pm 0.03$).
However, this good agreement  does not require the currently assumed metallicity dependence in the Cepheids PL relation. We postpone discussion of this issue to a companion paper.

There has been considerable discussion of the merits of the geometric distance to the maser sources in NGC 4258. It is seen in Figure \ref{final_calib} that our scale is only a 1$\sigma$ departure from the maser scale established by NGC 4258. However, the maser distance for this galaxy is more than a 2$\sigma$ departure from the TRGB measure.

Table \ref{tab_final_distances} provides a summary of the distance moduli determined for all the galaxies discussed in this paper. Our goal now is to apply the procedures described in Paper I and in this paper to the roughly 300 galaxies that have been observed appropriately by HST either with WFPC2 or ACS. The subsequent homogenous set of TRGB distances will provide a wealth of data on the distribution of galaxies and peculiar velocities within $\sim$ 7 Mpc.
\acknowledgements{This research has been supported by grants associated with Hubble Space Telescope programs AR-9950, GO-9771, GO-10210, and GO-10235}

\begin{deluxetable}{lccccc}
\tablewidth{0pt}
\tablecaption{Distance for all the galaxies discussed in this paper. \label{tab_final_distances}}
\tablehead{
\colhead{Galaxy} & \colhead{$I_{TRGB}$} & \colhead{$(V-I)_{TRGB}$} & \colhead{$A_I$} & \colhead{$M_I^{TRGB}$} & \colhead{$(m-M)_0$}}
\startdata
Sculptor &$15.60 \pm 0.03$ & $1.47 \pm 0.03$ & 0.03 & -4.08 & $19.64 \pm 0.04$ \\ 
Fornax &$16.75 \pm 0.03$ & $1.61 \pm 0.03$ & 0.04 & -4.05 & $20.76 \pm 0.04$ \\ 
NGC 185 &$20.35 \pm 0.02$ & $1.72 \pm 0.04$ & 0.36 & -4.02 & $24.01 \pm 0.04$ \\ 
NGC 224 &$20.53 \pm 0.07$ & $1.89 \pm 0.10$ & 0.15 & -3.99 & $24.37 \pm 0.08$ \\ 
IC1613 &$20.37 \pm 0.04$ & $1.56 \pm 0.02$ & 0.06 & -4.06 & $24.37 \pm 0.05$ \\ 
M33 (blue edge)\tablenotemark{1} &$20.73 \pm 0.02$ & $1.55 \pm 0.01$ & 0.08 & -4.06 & $24.71 \pm 0.03$ \\ 
M33 (red edge)\tablenotemark{1} &$20.91 \pm 0.02$ & $2.37 \pm 0.01$ & 0.08 & -3.88 & $24.71 \pm 0.04$ \\ 
WLM &$20.92 \pm 0.03$ & $1.47 \pm 0.02$ & 0.07 & -4.08 & $24.93 \pm 0.04$ \\ 
NGC 3109 &$21.62 \pm 0.04$ & $1.49 \pm 0.02$ & 0.13 & -4.07 & $25.56 \pm 0.05$ \\ 
Sextans A &$21.78 \pm 0.05$ & $1.37 \pm 0.04$ & 0.09 & -4.10 & $25.79 \pm 0.06$ \\ 
Sextans B &$21.76 \pm 0.03$ & $1.40 \pm 0.02$ & 0.06 & -4.09 & $25.79 \pm 0.04$ \\ 
NGC 300 &$22.54 \pm 0.02$ & $1.99 \pm 0.01$ & 0.03 & -3.97 & $26.48 \pm 0.04$ \\ 
NGC 3031 &$23.91 \pm 0.03$ & $2.09 \pm 0.02$ & 0.16 & -3.94 & $27.69 \pm 0.04$ \\ 
NGC 5128 &$24.03 \pm 0.02$ & $2.30 \pm 0.01$ & 0.21 & -3.90 & $27.72 \pm 0.04$ \\ 
NGC 1313 &$24.31 \pm 0.02$ & $1.60 \pm 0.00$ & 0.21 & -4.05 & $28.15 \pm 0.03$ \\ 
IC 4182 &$24.17 \pm 0.04$ & $1.41 \pm 0.01$ & 0.03 & -4.09 & $28.23 \pm 0.05$ \\ 
NGC 4605 &$24.70 \pm 0.03$ & $1.68 \pm 0.01$ & 0.03 & -4.03 & $28.71 \pm 0.04$ \\ 
NGC 3621 &$25.38 \pm 0.12$ & $1.65 \pm 0.03$ & 0.16 & -4.04 & $29.26 \pm 0.12$ \\ 
NGC 5457 &$25.31 \pm 0.08$ & $1.59 \pm 0.03$ & 0.02 & -4.05 & $29.34 \pm 0.09$ \\ 
NGC 4258 &$25.49 \pm 0.05$ & $2.03 \pm 0.02$ & 0.03 & -3.96 & $29.42 \pm 0.06$ \\ 
NGC 3351 &$25.92 \pm 0.04$ & $1.60 \pm 0.03$ & 0.05 & -4.05 & $29.92 \pm 0.05$ \\ 
\enddata
\tablenotetext{1}{We have used the wide color range of M33 RGB to independently derive distances from narrow color strips located at the red and blue edge of the RGB, respectively. Although the TRGB measurements are very different in the two cases, the final distance is consistent.}
\end{deluxetable}

\appendix
\section{CMD of the galaxies of the sample}
\label{cmds}
\begin{figure}
\epsscale{0.4}
\plotone{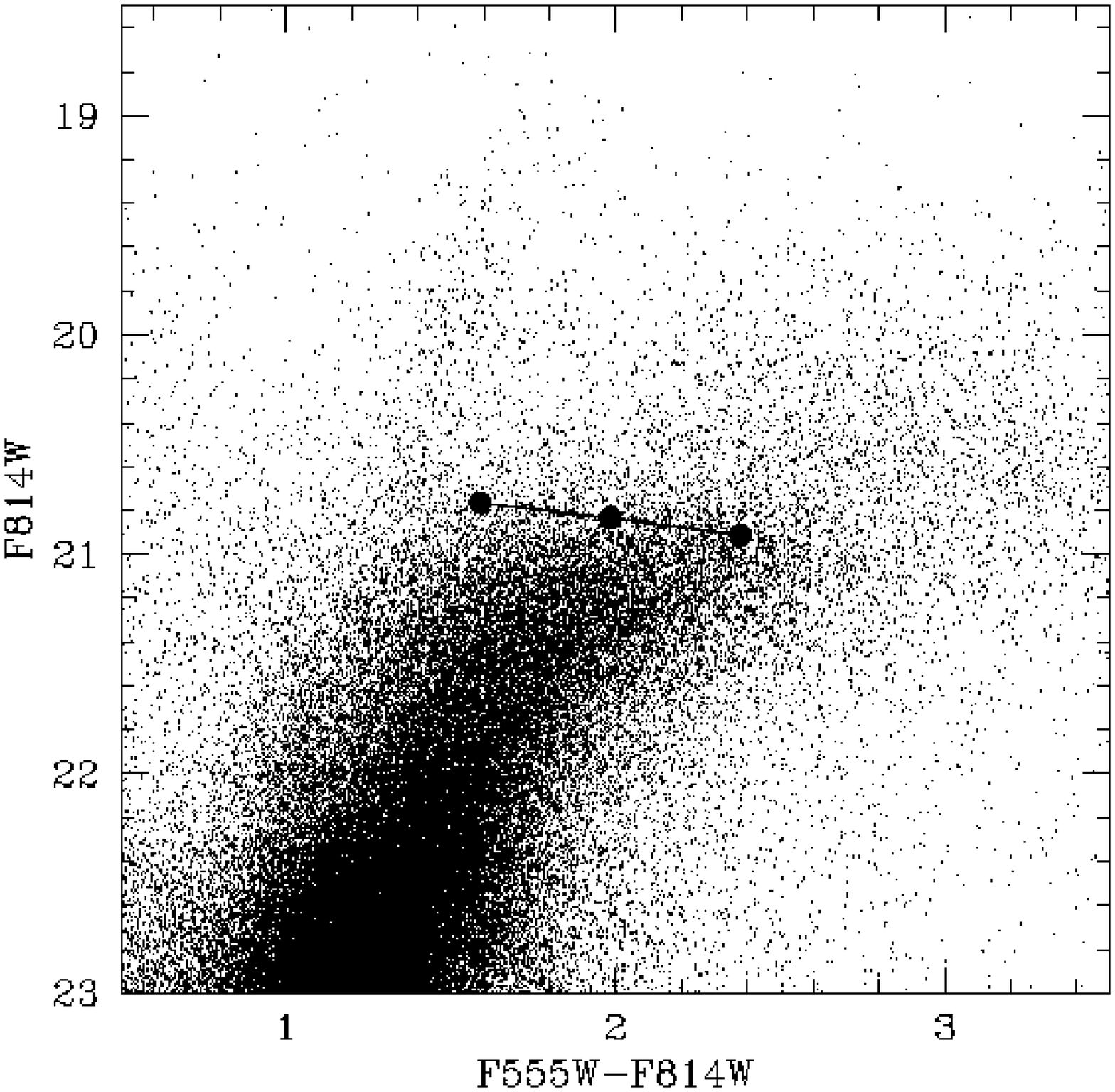}
\plotone{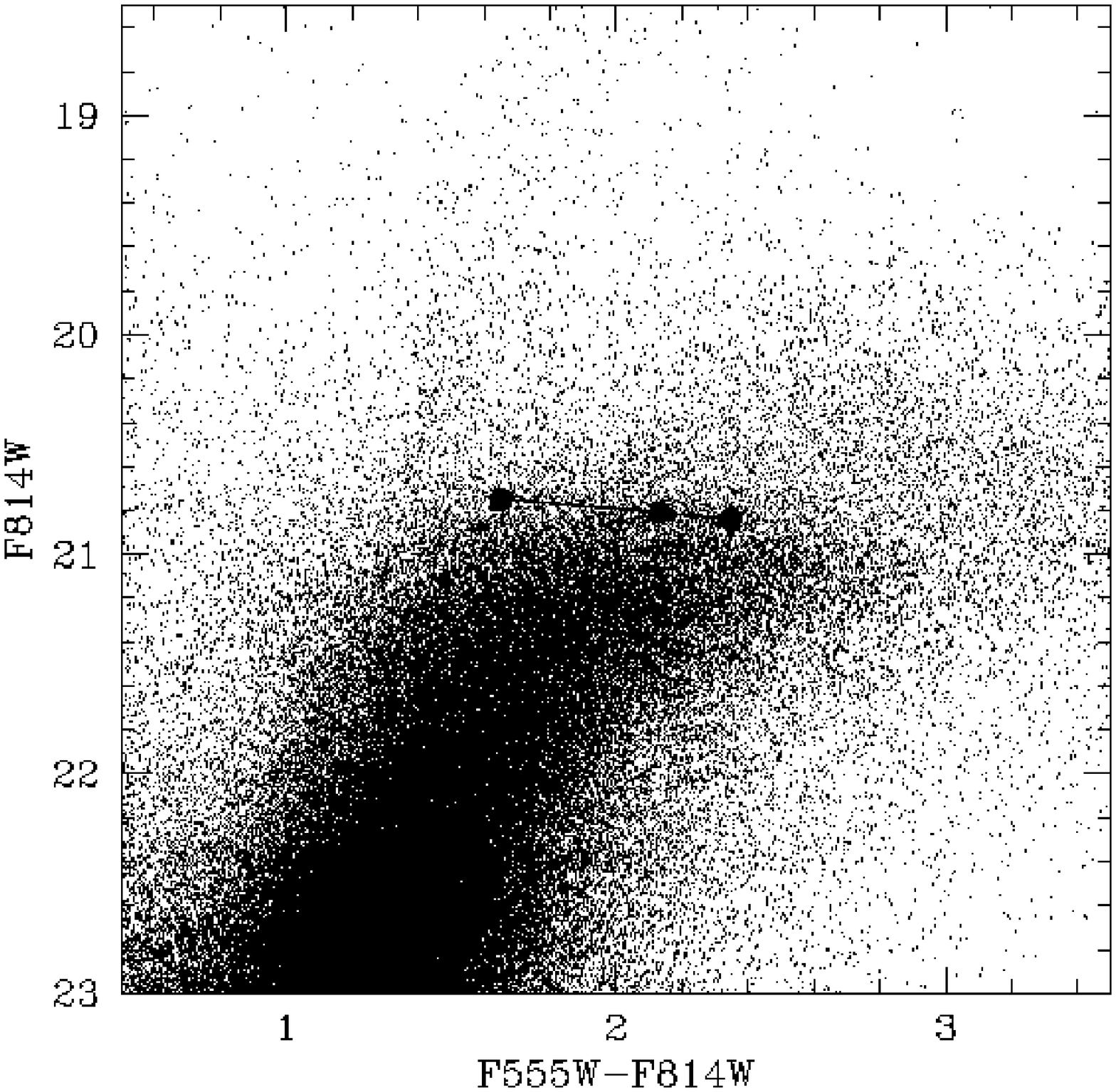}
\plotone{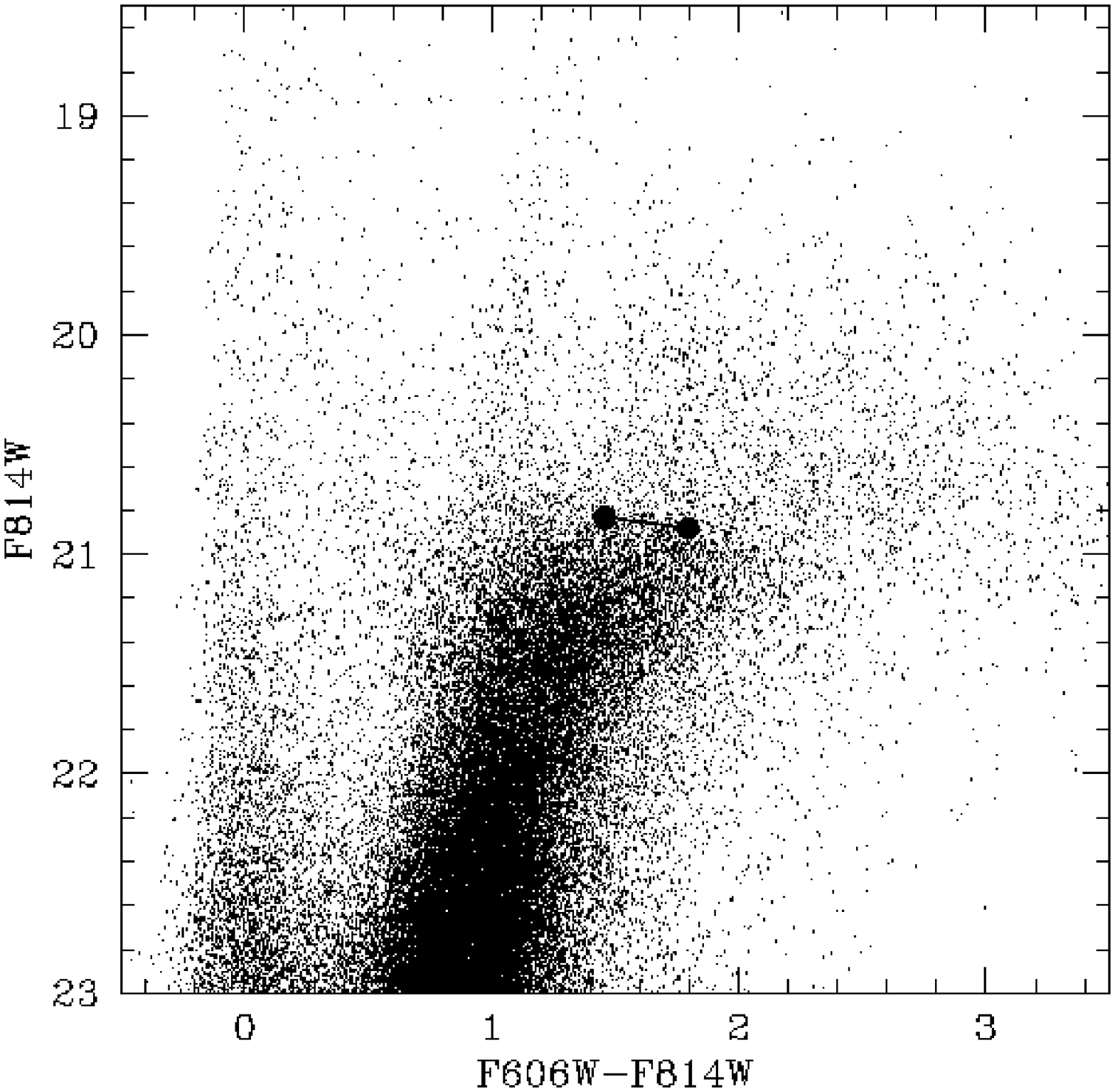}
\caption{Detection of the TRGB in M 33 for HST programs 5914, 6640, and 8059. \label{m33_1}}
\end{figure}

\begin{figure}
\epsscale{0.4}
\plotone{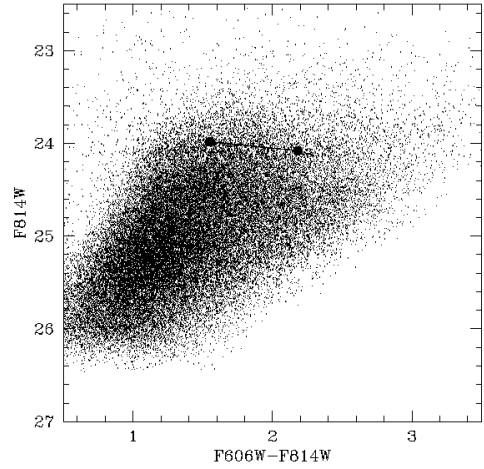}
\caption{Detection of the TRGB in NGC 5128 for HST program 8195 \label{ngc5128}}
\end{figure}

\begin{figure}
\epsscale{0.4}
\plotone{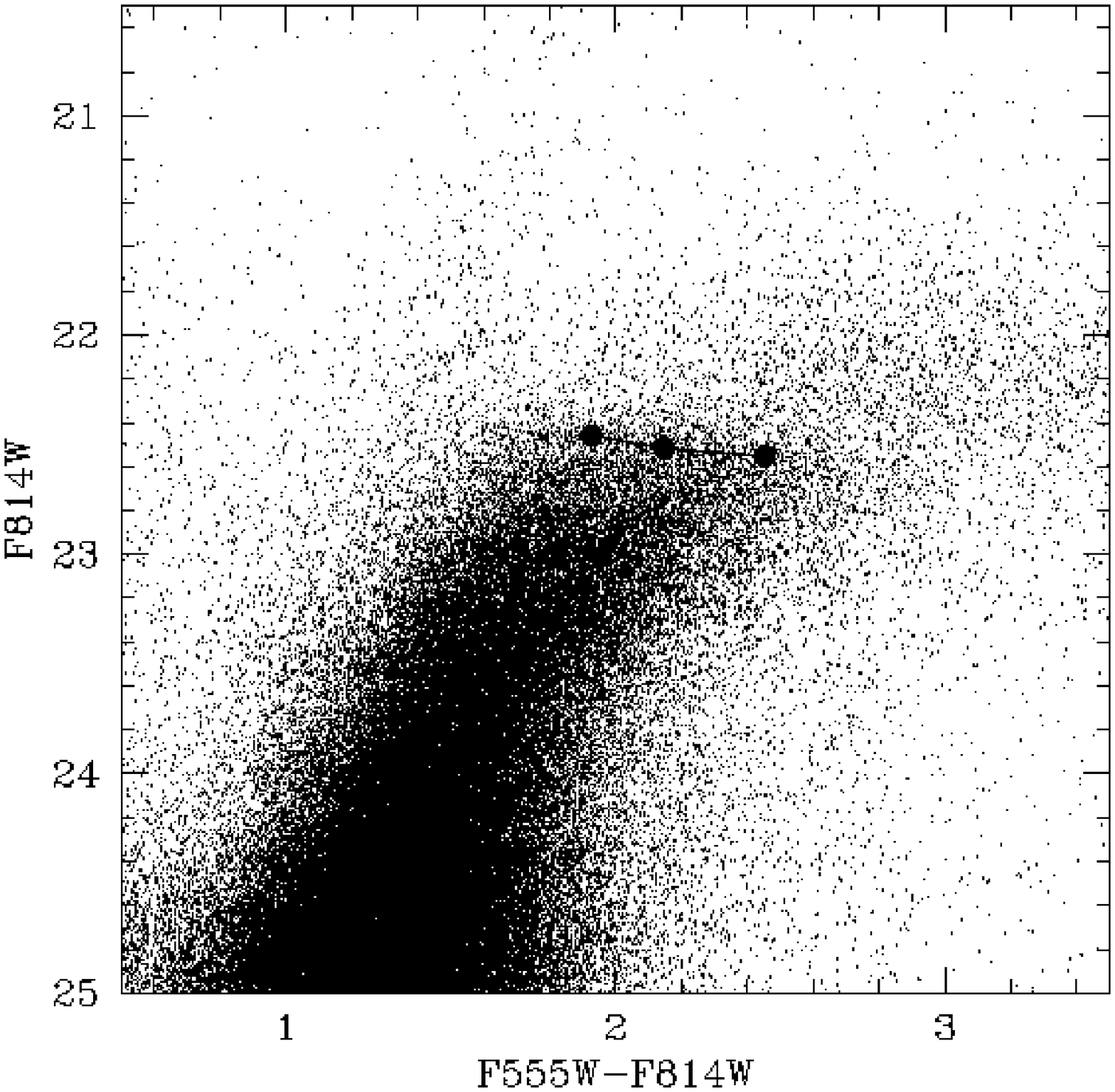}
\plotone{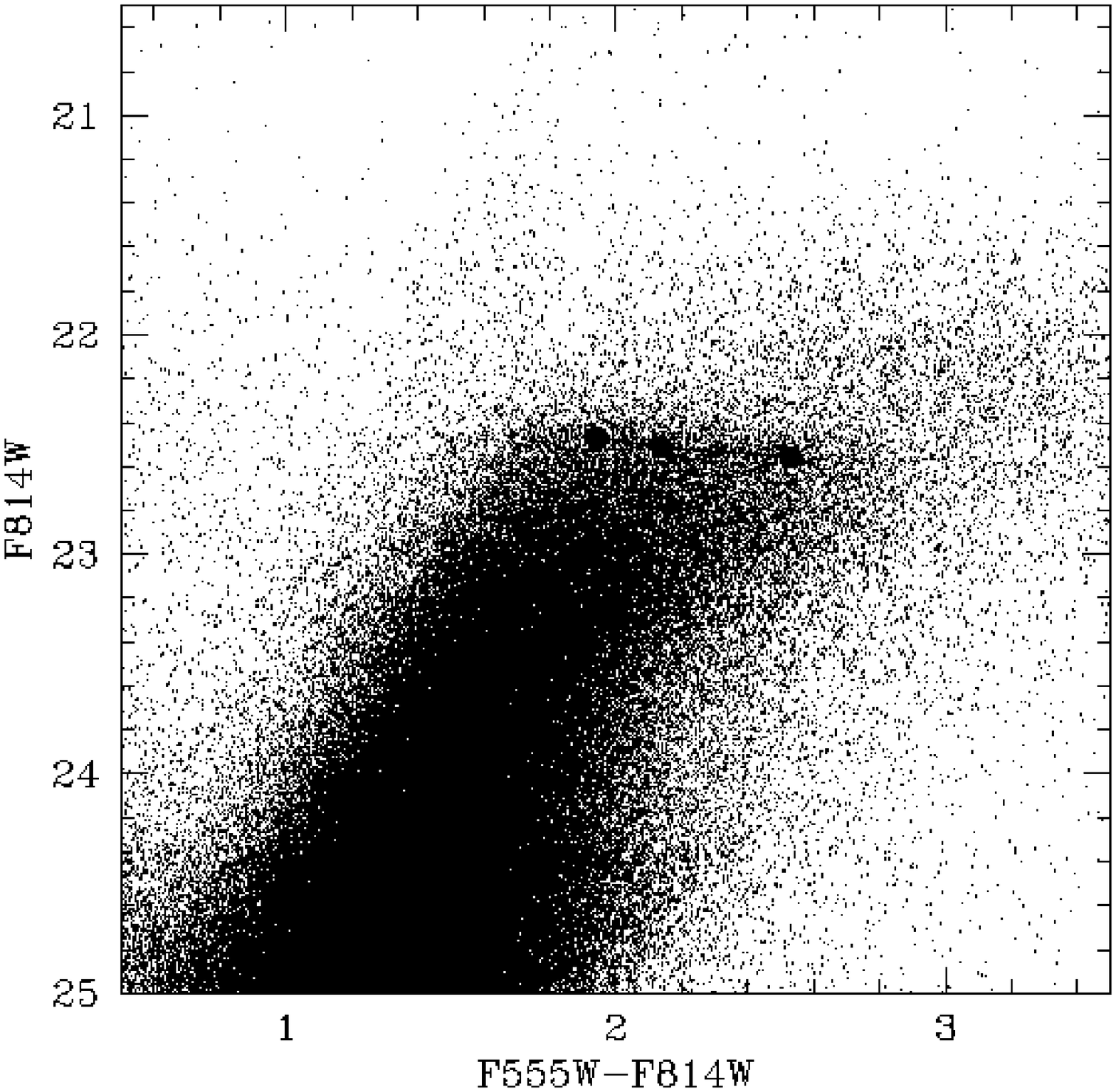}
\caption{Detection of the TRGB in NGC 300 for HST program 9492. Left panel presents the results for Field F2, right panel for field F3 as defined in \citet{Bresolin:rb}. \label{ngc300}}
\end{figure}

\begin{figure}
\epsscale{0.4}
\plotone{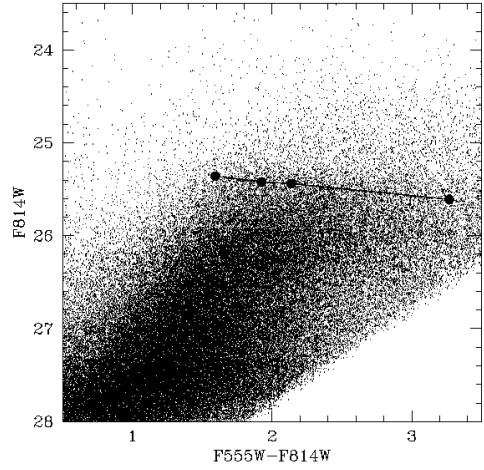}
\caption{Detection of the TRGB in NGC 4258 for HST program 9477.\label{ngc4258}}
\end{figure}

\begin{figure}
\epsscale{0.4}
\plotone{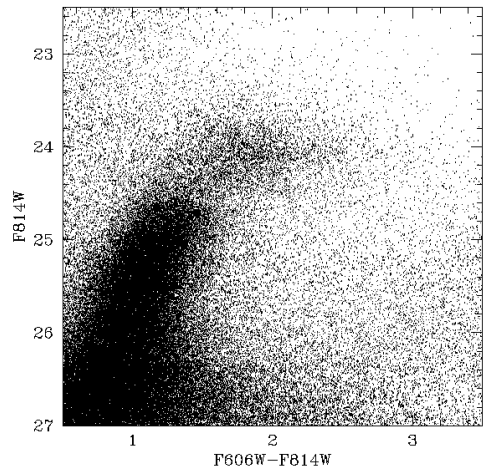}
\caption{Detection of the TRGB in NGC 4605 for HST program 9771. \label{ngc4605}}
\end{figure}

\begin{figure}
\epsscale{0.4}
\plotone{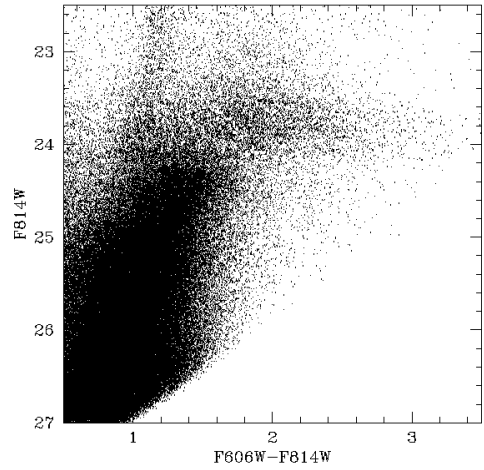}
\caption{Detection of the TRGB in NGC 1313 for HST program 10210.\label{ngc1313}}
\end{figure}


\begin{thebibliography}{55}
\expandafter\ifx\csname natexlab\endcsname\relax\def\natexlab#1{#1}\fi

\bibitem[{{Baade}(1944)}]{1944ApJ...100..137B}
{Baade}, W. 1944, \apj, 100, 137

\bibitem[{{Bellazzini} {et~al.}(2001){Bellazzini}, {Ferraro}, \&
  {Pancino}}]{2001ApJ...556..635B}
{Bellazzini}, M., {Ferraro}, F.~R., \& {Pancino}, E. 2001, \apj, 556, 635

\bibitem[{{Bellazzini} {et~al.}(2004){Bellazzini}, {Ferraro}, {Sollima},
  {Pancino}, \& {Origlia}}]{2004AA...424..199B}
{Bellazzini}, M., {Ferraro}, F.~R., {Sollima}, A., {Pancino}, E., \& {Origlia},
  L. 2004, \aap, 424, 199

\bibitem[{{Bonanos} {et~al.}(2006){Bonanos}, {Stanek}, {Kudritzki}, {Macri},
  {Sasselov}, {Kaluzny}, {Stetson}, {Bersier}, {Bresolin}, {Matheson},
  {Mochejska}, {Przybilla}, {Szentgyorgyi}, {Tonry}, \&
  {Torres}}]{2006ApJ...652..313B}
{Bonanos}, A.~Z. et al. 2006, \apj, 652, 313

\bibitem[{{Bresolin} {et~al.}(2005){Bresolin}, {Pietrzy{\'n}ski}, {Gieren}, \&
  {Kudritzki}}]{Bresolin:rb}
{Bresolin}, F., {Pietrzy{\'n}ski}, G., {Gieren}, W., \& {Kudritzki}, R.-P.
  2005, ApJ, in press

\bibitem[{{Butler} {et~al.}(2004){Butler}, {Mart{\'{\i}}nez-Delgado}, \&
  {Brandner}}]{2004AJ....127.1472B}
{Butler}, D.~J., {Mart{\'{\i}}nez-Delgado}, D., \& {Brandner}, W. 2004, \aj,
  127, 1472

\bibitem[{{Caputo} {et~al.}(2002){Caputo}, {Marconi}, \&
  {Musella}}]{2002ApJ...566..833C}
{Caputo}, F., {Marconi}, M., \& {Musella}, I. 2002, \apj, 566, 833

\bibitem[{{Carretta} \& {Gratton}(1997)}]{1997AAS..121...95C}
{Carretta}, E. \& {Gratton}, R.~G. 1997, \aaps, 121, 95

\bibitem[{{Carretta} {et~al.}(2000){Carretta}, {Gratton}, {Clementini}, \&
  {Fusi Pecci}}]{2000ApJ...533..215C}
{Carretta}, E., {Gratton}, R.~G., {Clementini}, G., \& {Fusi Pecci}, F. 2000,
  \apj, 533, 215

\bibitem[{{Cioni} {et~al.}(2006){Cioni}, {Girardi}, {Marigo}, \&
  {Habing}}]{2006AA...452..195C}
{Cioni}, M.-R.~L., {Girardi}, L., {Marigo}, P., \& {Habing}, H.~J. 2006, \aap,
  452, 195

\bibitem[{{Da Costa} \& {Armandroff}(1990)}]{1990AJ....100..162D}
{Da Costa}, G.~S. \& {Armandroff}, T.~E. 1990, \aj, 100, 162

\bibitem[{{Dolphin}(2000{\natexlab{a}})}]{2000PASP..112.1397D}
{Dolphin}, A.~E. 2000{\natexlab{a}}, \pasp, 112, 1397

\bibitem[{{Dolphin}(2000{\natexlab{b}})}]{2000PASP..112.1383D}
---. 2000{\natexlab{b}}, \pasp, 112, 1383

\bibitem[{{Ferrarese} {et~al.}(2006){Ferrarese}, {Mould}, {Stetson}, {Tonry},
  {Blakeslee}, \& {Ajhar}}]{2006astro.ph..5707F}
{Ferrarese}, L., {Mould}, J.~R., {Stetson}, P.~B., {Tonry}, J.~L., {Blakeslee},
  J.~P., \& {Ajhar}, E.~A. 2006, ArXiv Astrophysics e-prints

\bibitem[{{Ferraro} {et~al.}(1999){Ferraro}, {Messineo}, {Fusi Pecci}, {de
  Palo}, {Straniero}, {Chieffi}, \& {Limongi}}]{1999AJ....118.1738F}
{Ferraro}, F.~R., {Messineo}, M., {Fusi Pecci}, F., {de Palo}, M.~A.,
  {Straniero}, O., {Chieffi}, A., \& {Limongi}, M. 1999, \aj, 118, 1738

\bibitem[{{Ferraro} {et~al.}(2000){Ferraro}, {Montegriffo}, {Origlia}, \& {Fusi
  Pecci}}]{2000AJ....119.1282F}
{Ferraro}, F.~R., {Montegriffo}, P., {Origlia}, L., \& {Fusi Pecci}, F. 2000,
  \aj, 119, 1282

\bibitem[{{Freedman} {et~al.}(2001){Freedman}, {Madore}, {Gibson}, {Ferrarese},
  {Kelson}, {Sakai}, {Mould}, {Kennicutt}, {Ford}, {Graham}, {Huchra},
  {Hughes}, {Illingworth}, {Macri}, \& {Stetson}}]{2001ApJ...553...47F}
{Freedman}, W.~L. et al. 2001, \apj, 553, 47

\bibitem[{{Freedman} {et~al.}(1992){Freedman}, {Madore}, {Hawley}, {Horowitz},
  {Mould}, {Navarrete}, \& {Sallmen}}]{1992ApJ...396...80F}
{Freedman}, W.~L., {Madore}, B.~F., {Hawley}, S.~L., {Horowitz}, I.~K.,
  {Mould}, J., {Navarrete}, M., \& {Sallmen}, S. 1992, \apj, 396, 80

\bibitem[{{Gieren} {et~al.}(2005){Gieren}, {Pietrzy{\'n}ski}, {Soszy{\'n}ski},
  {Bresolin}, {Kudritzki}, {Minniti}, \& {Storm}}]{2005ApJ...628..695G}
{Gieren}, W., {Pietrzy{\'n}ski}, G., {Soszy{\'n}ski}, I., {Bresolin}, F.,
  {Kudritzki}, R.-P., {Minniti}, D., \& {Storm}, J. 2005, \apj, 628, 695

\bibitem[{{Girardi} {et~al.}(2002){Girardi}, {Bertelli}, {Bressan}, {Chiosi},
  {Groenewegen}, {Marigo}, {Salasnich}, \& {Weiss}}]{2002AA...391..195G}
{Girardi}, L., {Bertelli}, G., {Bressan}, A., {Chiosi}, C., {Groenewegen},
  M.~A.~T., {Marigo}, P., {Salasnich}, B., \& {Weiss}, A. 2002, \aap, 391, 195

\bibitem[{{Harris} {et~al.}(1999){Harris}, {Harris}, \&
  {Poole}}]{1999AJ....117..855H}
{Harris}, G.~L.~H., {Harris}, W.~E., \& {Poole}, G.~B. 1999, \aj, 117, 855

\bibitem[{{Herrnstein} {et~al.}(1999){Herrnstein}, {Moran}, {Greenhill},
  {Diamond}, {Inoue}, {Nakai}, {Miyoshi}, {Henkel}, \&
  {Riess}}]{1999Natur.400..539H}
{Herrnstein}, J.~R. et al. 1999,
  \nat, 400, 539

\bibitem[{{Holtzman} {et~al.}(1995){Holtzman}, {Burrows}, {Casertano},
  {Hester}, {Trauger}, {Watson}, \& {Worthey}}]{1995PASP..107.1065H}
{Holtzman}, J.~A., {Burrows}, C.~J., {Casertano}, S., {Hester}, J.~J.,
  {Trauger}, J.~T., {Watson}, A.~M., \& {Worthey}, G. 1995, \pasp, 107, 1065

\bibitem[{{Humphreys} {et~al.}(2005){Humphreys}, {Argon}, {Greenhill}, {Moran},
  \& {Reid}}]{2005ASPC..340..466H}
{Humphreys}, E.~M.~L., {Argon}, A.~L., {Greenhill}, L.~J., {Moran}, J.~M., \&
  {Reid}, M.~J. 2005, in ASP Conf. Ser. 340: Future Directions in High
  Resolution Astronomy, ed. J.~{Romney} \& M.~{Reid}, 466--+

\bibitem[{{Karachentsev} {et~al.}(2006){Karachentsev}, {Dolphin}, {Tully},
  {Sharina}, {Makarova}, {Makarov}, {Karachentseva}, {Sakai}, \&
  {Shaya}}]{2006AJ....131.1361K}
{Karachentsev}, I.~D. et al. 2006, \aj, 131, 1361

\bibitem[{{Kim} {et~al.}(2002){Kim}, {Kim}, {Lee}, {Sarajedini}, \&
  {Geisler}}]{2002AJ....123..244K}
{Kim}, M., {Kim}, E., {Lee}, M.~G., {Sarajedini}, A., \& {Geisler}, D. 2002,
  \aj, 123, 244

\bibitem[{{Lee} {et~al.}(1993){Lee}, {Freedman}, \&
  {Madore}}]{1993ApJ...417..553L}
{Lee}, M.~G., {Freedman}, W.~L., \& {Madore}, B.~F. 1993, \apj, 417, 553

\bibitem[{{Lee} {et~al.}(2002){Lee}, {Kim}, {Sarajedini}, {Geisler}, \&
  {Gieren}}]{2002ApJ...565..959L}
{Lee}, M.~G., {Kim}, M., {Sarajedini}, A., {Geisler}, D., \& {Gieren}, W. 2002,
  \apj, 565, 959

\bibitem[{{Lee} {et~al.}(1990){Lee}, {Demarque}, \&
  {Zinn}}]{1990ApJ...350..155L}
{Lee}, Y.-W., {Demarque}, P., \& {Zinn}, R. 1990, \apj, 350, 155

\bibitem[{{Macri} {et~al.}(2006){Macri}, {Stanek}, {Bersier}, {Greenhill}, \&
  {Reid}}]{2006astro.ph..8211M}
{Macri}, L.~M., {Stanek}, K.~Z., {Bersier}, D., {Greenhill}, L.~J., \& {Reid},
  M.~J. 2006, \apj, 652, 1133

\bibitem[{{Madore} \& {Freedman}(1991)}]{1991PASP..103..933M}
{Madore}, B.~F. \& {Freedman}, W.~L. 1991, \pasp, 103, 933

\bibitem[{{Madore} \& {Freedman}(1995)}]{1995AJ....109.1645M}
---. 1995, \aj, 109, 1645

\bibitem[{{Madore} {et~al.}(1997){Madore}, {Freedman}, \&
  {Sakai}}]{1997eds..proc..239M}
{Madore}, B.~F., {Freedman}, W.~L., \& {Sakai}, S. 1997, in The Extragalactic
  Distance Scale, ed. M.~{Livio}, M.~{Donahue}, \& N.~{Panagia}, 239--253

\bibitem[{{Madore} {et~al.}(1987){Madore}, {Welch}, {McAlary}, \&
  {McLaren}}]{1987ApJ...320...26M}
{Madore}, B.~F., {Welch}, D.~L., {McAlary}, C.~W., \& {McLaren}, R.~A. 1987,
  \apj, 320, 26

\bibitem[{{Magrini} {et~al.}(2000){Magrini}, {Corradi}, {Mampaso}, \&
  {Perinotto}}]{2000AA...355..713M}
{Magrini}, L., {Corradi}, R.~L.~M., {Mampaso}, A., \& {Perinotto}, M. 2000,
  \aap, 355, 713

\bibitem[{{Makarov} {et~al.}(2006){Makarov}, {Makarova}, {Rizzi}, {Tully},
  {Dolphin}, {Sakai}, \& {Shaya}}]{2006AJ....132.2729M}
{Makarov}, D., {Makarova}, L., {Rizzi}, L., {Tully}, R.~B., {Dolphin}, A.~E.,
  {Sakai}, S., \& {Shaya}, E.~J. 2006, \aj, 132, 2729 (Paper I)

\bibitem[{{M{\'e}ndez} {et~al.}(2002){M{\'e}ndez}, {Davis}, {Moustakas},
  {Newman}, {Madore}, \& {Freedman}}]{2002AJ....124..213M}
{M{\'e}ndez}, B., {Davis}, M., {Moustakas}, J., {Newman}, J., {Madore}, B.~F.,
  \& {Freedman}, W.~L. 2002, \aj, 124, 213

\bibitem[{{Newman} {et~al.}(2001){Newman}, {Ferrarese}, {Stetson}, {Maoz},
  {Zepf}, {Davis}, {Freedman}, \& {Madore}}]{2001ApJ...553..562N}
{Newman}, J.~A., {Ferrarese}, L., {Stetson}, P.~B., {Maoz}, E., {Zepf}, S.~E.,
  {Davis}, M., {Freedman}, W.~L., \& {Madore}, B.~F. 2001, \apj, 553, 562

\bibitem[{{Pierce} {et~al.}(2000){Pierce}, {Jurcevic}, \&
  {Crabtree}}]{2000MNRAS.313..271P}
{Pierce}, M.~J., {Jurcevic}, J.~S., \& {Crabtree}, D. 2000, \mnras, 313, 271

\bibitem[{{Rejkuba}(2004)}]{2004AA...413..903R}
{Rejkuba}, M. 2004, \aap, 413, 903

\bibitem[{{Rizzi} {et~al.}(2006){Rizzi}, {Bresolin}, {Kudritzki}, {Gieren}, \&
  {Pietrzy{\'n}ski}}]{2006ApJ...638..766R}
{Rizzi}, L., {Bresolin}, F., {Kudritzki}, R.-P., {Gieren}, W., \&
  {Pietrzy{\'n}ski}, G. 2006, \apj, 638, 766

\bibitem[{{Sakai} {et~al.}(2004){Sakai}, {Ferrarese}, {Kennicutt}, \&
  {Saha}}]{2004ApJ...608...42S}
{Sakai}, S., {Ferrarese}, L., {Kennicutt}, R.~C., \& {Saha}, A. 2004, \apj,
  608, 42

\bibitem[{{Sakai} {et~al.}(1996){Sakai}, {Madore}, \&
  {Freedman}}]{1996ApJ...461..713S}
{Sakai}, S., {Madore}, B.~F., \& {Freedman}, W.~L. 1996, \apj, 461, 713

\bibitem[{{Salaris} \& {Cassisi}(1997)}]{1997MNRAS.289..406S}
{Salaris}, M. \& {Cassisi}, S. 1997, \mnras, 289, 406

\bibitem[{{Salaris} \& {Cassisi}(1998)}]{1998MNRAS.298..166S}
---. 1998, \mnras, 298, 166

\bibitem[{{Salaris} {et~al.}(2002){Salaris}, {Cassisi}, \&
  {Weiss}}]{2002PASP..114..375S}
{Salaris}, M., {Cassisi}, S., \& {Weiss}, A. 2002, \pasp, 114, 375

\bibitem[{{Salasnich} {et~al.}(2000){Salasnich}, {Girardi}, {Weiss}, \&
  {Chiosi}}]{2000AA...361.1023S}
{Salasnich}, B., {Girardi}, L., {Weiss}, A., \& {Chiosi}, C. 2000, \aap, 361,
  1023

\bibitem[{{Sarajedini} {et~al.}(2006){Sarajedini}, {Barker}, {Geisler},
  {Harding}, \& {Schommer}}]{2006AJ....132.1361S}
{Sarajedini}, A., {Barker}, M.~K., {Geisler}, D., {Harding}, P., \& {Schommer},
  R. 2006, \aj, 132, 1361

\bibitem[{{Sarajedini} {et~al.}(2000){Sarajedini}, {Geisler}, {Schommer}, \&
  {Harding}}]{2000AJ....120.2437S}
{Sarajedini}, A., {Geisler}, D., {Schommer}, R., \& {Harding}, P. 2000, \aj,
  120, 2437

\bibitem[{{Schlegel} {et~al.}(1998){Schlegel}, {Finkbeiner}, \&
  {Davis}}]{1998ApJ...500..525S}
{Schlegel}, D.~J., {Finkbeiner}, D.~P., \& {Davis}, M. 1998, \apj, 500, 525

\bibitem[{{Sirianni} {et~al.}(2005){Sirianni}, {Jee}, {Ben{\'{\i}}tez},
  {Blakeslee}, {Martel}, {Meurer}, {Clampin}, {De Marchi}, {Ford}, {Gilliland},
  {Hartig}, {Illingworth}, {Mack}, \& {McCann}}]{2005PASP..117.1049S}
{Sirianni}, M. et al. 2005, \pasp, 117, 1049

\bibitem[{{Tikhonov} {et~al.}(2005){Tikhonov}, {Galazutdinova}, \&
  {Drozdovsky}}]{2005AA...431..127T}
{Tikhonov}, N.~A., {Galazutdinova}, O.~A., \& {Drozdovsky}, I.~O. 2005, \aap,
  431, 127

\bibitem[{{Tonry} {et~al.}(2001){Tonry}, {Dressler}, {Blakeslee}, {Ajhar},
  {Fletcher}, {Luppino}, {Metzger}, \& {Moore}}]{2001ApJ...546..681T}
{Tonry}, J.~L., {Dressler}, A., {Blakeslee}, J.~P., {Ajhar}, E.~A., {Fletcher},
  A.~B., {Luppino}, G.~A., {Metzger}, M.~R., \& {Moore}, C.~B. 2001, \apj, 546,
  681

\bibitem[{{Walker}(1988)}]{1988PASP..100..949W}
{Walker}, A.~R. 1988, \pasp, 100, 949

\bibitem[{{Zinn} \& {West}(1984)}]{1984ApJS...55...45Z}
{Zinn}, R. \& {West}, M.~J. 1984, \apjs, 55, 45

\end{thebibliography}
\end{document}